\documentclass{article}
\usepackage{colortbl}
\usepackage[english]{babel}
\usepackage{amsmath,mathrsfs,amssymb}
\usepackage{etoolbox}
\usepackage{hyperref}
\usepackage{amsfonts}
\usepackage{amssymb}
\usepackage{scalerel}
\usepackage{dsfont}	
\usepackage{graphicx}
\usepackage{bbm}
\usepackage{pdflscape}
\usepackage{afterpage}
\usepackage{cite}
\usepackage{footnote}
\usepackage{datatool}
\usepackage{pdfpages}
\usepackage{subfigure}
\usepackage{mathtools}
\usepackage{acronym}
\usepackage{extarrows}
\usepackage{tikz-cd} % Required for inserting images

\def\hybrid{
        \topmargin -20pt
        \oddsidemargin 0pt
        \headheight 0pt \headsep 0pt
        \textwidth 6.25in % A4 paper
        \textheight 9.5in % A4 paper
        \marginparwidth .875in
        \parskip 5pt plus 1pt \jot = 1.5ex}

\def\del{\partial}
\def\BVB{{\rm BV}_\infty^\square}
\def\BV{{\rm BV}^\square}

\def\cL{{\cal L}}

\def\cF{{\cal F}}

\def\cA{{\cal A}}
\def\cE{{\cal E}}

\def\cN{{\cal N}}

\def\cV{{\cal V}}

\def\cK{{\cal K}}
\def\cX{{\cal X}}

\def\d{\scaleobj{1.1}{{\mathbbm{d}}}}
\def\B{\square}

\hybrid

 \csname
@addtoreset\endcsname{equation}{section}

\begin{document}

\begin{titlepage}
\rightline{}
%\rightline\today
\rightline{June 2023}
\rightline{HU-EP-23/17-RTG}  
\begin{center}
\vskip 1.5cm
{\Large \bf{Gauge independent kinematic algebra of self-dual Yang-Mills theory}}
\vskip 1.5cm

{\large\bf {Roberto Bonezzi$^{\dagger}$, Felipe D\'iaz-Jaramillo$^{\dagger}$, Silvia Nagy$^*$
}}

\vskip 1cm

{\it $^\dagger$ Institute for Physics, Humboldt University Berlin,\\
 Zum Gro\ss en Windkanal 6, D-12489 Berlin, Germany}\\[2 ex] 

 {\it $^*$ Department of Mathematical Sciences, Durham University, Durham, DH1 3LE, UK}\\[2 ex] 
 
roberto.bonezzi@physik.hu-berlin.de, 
felipe.diaz-jaramillo@hu-berlin.de, silvia.nagy@durham.ac.uk,
\vskip .1cm

\vskip .2cm

\end{center}

\bigskip\bigskip
\begin{center} 
\textbf{Abstract}

\end{center} 
\begin{quote}

The double copy programme relies crucially on the so-called color-kinematics duality which, in turn, is widely believed to descend from a kinematic algebra possessed by gauge theories.
In this paper we construct the kinematic algebra of gauge invariant and off-shell self-dual Yang-Mills theory, up to trilinear maps. This structure is a homotopy algebra of the same type as the ones recently uncovered in Chern-Simons and full Yang-Mills theories.
To make contact with known results for the self-dual sector, we show that it reduces to the algebra found by Monteiro and O'Connell upon taking light-cone gauge and partially solving the self-duality constraints.
Finally, we test a double copy prescription recently proposed in \cite{Bonezzi:2022bse} and reproduce self-dual gravity.

\end{quote} 
\vfill
\setcounter{footnote}{0}
\end{titlepage}

\tableofcontents
\newpage
\section{Introduction}

Color-kinematics duality \cite{Bern:2008qj,Bern:2010ue} is a remarkable insight into the structure of gauge theory. It underpins much of the double copy programme\footnote{See \cite{Bern:2019prr,Adamo:2022dcm} and references within.}, which aims to construct gravitational quantities from two copies of their Yang-Mills (YM) counterparts. The duality was originally discovered in the context of scattering amplitudes in gauge theory \cite{Bern:2008qj,Bern:2010ue,Stieberger:2009hq,Bjerrum-Bohr:2009ulz,Feng:2010my} where the so-called kinematic factors (roughly speaking the part of the amplitudes depending on polarisation vectors and momenta) were found to satisfy the same relations as the color factors (given by the structure constants of the non-abelian gauge group). In particular, both the kinematic and the color factors were found to satisfy Jacobi-like relations. In the latter, these follow naturally from the underlying gauge group. This suggested the existence of an additional, previously unexplored symmetry algebra in YM, dubbed the kinematic algebra\cite{Monteiro:2011pc,Chen:2019ywi,Chen:2022nei,Chen:2021chy,Brandhuber:2022enp,Brandhuber:2021bsf,Ben-Shahar:2021zww,Mafra:2014oia,Mafra:2015vca,Lee:2015upy,Ben-Shahar:2021doh,Borsten:2023reb}.

In this article we will focus on the self-dual sector of YM/gravity. This is a subsector obtained via a constraint on the field strength/ Riemann curvature. Alternatively, from a particle perspective, it describes only one of the two helicities of the theory. It has been extensively studied, due to its integrability \cite{Ward:1977ta,Dunajski_1998,Prasad:1979zc,Dolan:1983bp,Popov:1996uu,Popov:1998pc,Park:1989vq,Husain:1993dp}, its role in the construction of instantons\footnote{Also shown to be connected via the double copy \cite{Berman:2018hwd,Armstrong-Williams:2022apo,Luna:2018dpt
}.} \cite{Belavin:1975fg,Atiyah:1978ri,Eguchi:1976db,Belavin:1976rx,Eguchi:1980jx}, and the special part it plays in the context of scattering amplitudes. In this latter case, it has been shown that the only non-trivial contribution comes in at 1 loop \cite{Bern:1993qk,Mahlon:1993si,Bern:1996ja,Bern:1998xc,Bern:1998sv
}, computed to all orders in \cite
{Chattopadhyay:2020oxe,Chattopadhyay:2021udc}; via unitarity cuts, this was shown to control the divergence of quantum gravity at two loops \cite{Bern:2017puu}. Another recent application is in the context of celestial holography, where infinite towers of soft symmetries are organised into the $w_{1+\infty}$ algebra\footnote{More precisely, the loop algebra of the wedge subalgebra of $w_{1+\infty}$.} \cite{Guevara:2021abz,Strominger:2021lvk,Himwich:2021dau,Jiang:2021ovh}, which turns out to have a natural action in this sector \cite{Boyer:1985aj,Park:1989fz,Park:1989vq,Adamo:2021lrv,Monteiro:2022lwm}

The self-dual sector was the setting of the first explicit realisation of the aforementioned kinematic algebra, specifically in the light-cone gauge \cite{Monteiro:2011pc}. This led to an elegant double copy to self-dual gravity, further explored in \cite{Chacon:2020fmr,Campiglia:2021srh,Monteiro:2022nqt,Monteiro:2013rya,Cheung:2016prv,Chen:2019ywi,Elor:2020nqe,Armstrong-Williams:2022apo,Farnsworth:2021wvs,Skvortsov:2022unu,He:2015wgf,Berman:2018hwd,Nagy:2022xxs,Monteiro:2022lwm,Krasnov:2021cva,Monteiro:2022xwq,Lipstein:2023pih}, with a description of the kinematic algebra recently given via its relation to twistor theory \cite{Borsten:2022vtg}. The self-dual sector is more than just a convenient toy model; it is in fact possible to recast the full YM/gravity theory as a perturbation around this sector \cite{Chalmers:1996rq,Abou-Zeid:2005zfo}. This is different from the usual notion of perturbation theory - schematically it consists of reintroducing the missing helicity field in a controlled manner. Here initial steps have already been taken in extending the kinematic algebra \cite{Chen:2019ywi,Chen:2021chy,Ben-Shahar:2022ixa}.

A crucial element of the above constructions is making an appropriate gauge choice. In light-cone gauge, this leads to a description of the self-dual fields in terms of scalars \cite{Plebanski:1975wn,Bardeen:1995gk,Chalmers:1996rq,Prasad:1979zc,Dolan:1983bp,Parkes:1992rz,Cangemi:1996pf,Popov:1996uu,Popov:1998pc}. In this context, the kinematic algebra emerges naturally as an area-preserving diffeomorphism algebra \cite{Monteiro:2011pc}, whose Poisson bracket is manifest in the self-duality equation governing the dynamics of the system. We note that this reflects a more general feature of the double copy, both in the construction of amplitudes, and classical solutions: it is often the case that a judicious choice of gauge (or field redefinition) facilitates the map between YM and gravity.

It then follows that one of the key questions on the formal side of the double copy is to understand the extent to which the programme relies on these specific gauge choices. For the case at hand: does the self-dual kinematic algebra survive in the absence of gauge fixing? We address this question via the language of homotopy algebras, which has given some encouraging results in formulating the double copy in a gauge independent manner \cite{Bonezzi:2022bse,Bonezzi:2022yuh}, up to quartic order in perturbation theory. There exists a related proposal for an off-shell double copy construction, based on the BRST/BV formulation of the gauge theory and gravity, respectively \cite{Borsten:2021hua,Borsten:2021gyl,Borsten:2020zgj,Borsten:2021zir,Borsten:2020xbt,Godazgar:2022gfw,Luna:2020adi,Borsten:2019prq,Anastasiou:2018rdx}. This allows us to give an explicit off-shell description of the kinematic algebra, without the need to go to a particular gauge. We show that this is not a strict algebra, but instead a BV$_{\infty}^{\square}$ algebra.
Despite the exotic name, this is an encouraging result: the same type of algebras have recently been proposed as the underlying kinematic algebras of off-shell Chern-Simons and Yang-Mills theories \cite{Reiterer:2019dys,Bonezzi:2022bse,Borsten:2022vtg}. Our construction thus strengthens the claim that this kind of algebraic structure, which is ultimately responsible for the consistency of the double copy, is present in gauge theories beyond the context of scattering amplitudes. To relate our findings with the known kinematic algebra of the self-dual sector, we demonstrate how this reduces to the established results upon gauge fixing and partially solving the self-duality equations, and we also give the double copy formulation in the language of $L_{\infty}$ algebras, adapting the procedure proposed in \cite{Bonezzi:2022bse,Bonezzi:2023ced}.

The article is organised as follows. In \autoref{Sec:Loo} we give a brief description of the self-dual sector of YM, and introduce the concept of $L_{\infty}$ algebras, with a particular focus on reformulating self-dual Yang-Mills (SDYM) in this framework. In \autoref{Sec:BVoo} we construct the kinematic algebra without resorting to gauge fixing. We identify this as a BV$_{\infty}^{\square}$ algebra, and give an explicit construction up to trilinear maps. Moreover, we illustrate the gauge independent double copy prescription of \cite{Bonezzi:2022bse, Bonezzi:2022yuh}, and construct linearised self-dual gravity. Then, in \autoref{Light-cone gauge and area preserving diffeomorphisms}, we show how this reduces to a BV$^{\square}$ algebra containing the expected area-preserving diffeomorphisms, upon going to light-cone gauge and partially solving the self-duality equations. Interestingly, we find that this is already present before going to the scalar description of the self-dual sector. \autoref{Double Copy and Self-Dual Gravity} is dedicated to applying the double copy algorithm developed in \cite{Bonezzi:2022bse,Bonezzi:2022yuh} to our model in light-cone gauge. We find that we reproduce the Plebanski description of self-dual gravity to all orders, thus making contact with the seminal results of \cite{Monteiro:2011pc}. We conclude and discuss possible applications in \autoref{Conclusions and Outlook}.

\section{$L_{\infty}$ algebras and self-dual Yang-Mills}\label{Sec:Loo}

In this section we present the $L_\infty$ algebra
of self-dual Yang-Mills theory, which will serve as the starting point to studying its kinematic algebra in the next section. We start by briefly reviewing self-dual Yang-Mills theory, where we set up our notation and conventions and proceed with a brief introduction to $L_\infty$ algebras, before specializing to the $L_\infty$ algebra of SDYM. 

\subsection{Self-dual Yang-Mills}

Here we work on four dimensional flat Euclidean space with coordinates $x^\mu=(t,x,y,z)$, where $t$ is Euclidean time, and flat metric $\delta_{\mu\nu}$ in the cartesian coordinates $x^\mu$. In self-dual Yang-Mills the field strength $F_{\mu\nu}$ is subject to the self-duality constraint
\begin{equation}\label{componentSD}
F_{\mu\nu}=\frac12\,\epsilon_{\mu\nu\rho\sigma}\,F^{\rho\sigma}\;,  
\end{equation}
which is stronger and implies the standard Yang-Mills equation $D^\mu F_{\mu\nu}=0$ upon taking a covariant divergence of \eqref{componentSD}.

In this and the next section we will use differential form language. We normalise $p-$forms in the standard way: $\omega_p=\frac{1}{p!}\,\omega_{\mu_1\cdots\mu_p}dx^{\mu_1}\wedge\cdots\wedge dx^{\mu_p}$, and define the Hodge star as
\begin{equation}\label{Hodge}
\star\,\omega_p:=\frac{1}{p!(4-p)!}\,\epsilon_{\mu_1\cdots\mu_{4-p}\nu_1\cdots\nu_p}\omega^{\nu_1\cdots\nu_p}\,dx^{\mu_1}\wedge\cdots\wedge dx^{\mu_{4-p}}\;.
\end{equation}
With this definition, the self-duality constraint becomes $F=\star F$. The Hodge star obeys $\star^2=(-1)^p$ in four dimensions, which allows us to define the (anti)self-dual projectors on the space of two-forms:
\begin{equation}
P_\pm:=\tfrac12\,(1\pm\star)\;,\quad P_\pm^2=P_\pm\;,\quad P_\pm P_\mp=0\;.    
\end{equation}
The dynamical field in self-dual Yang-Mills theory is a Lie algebra-valued one-form $A=A_\mu^a\, dx^\mu\,T_a$, where $T_{a}$ are generators of the color Lie algebra $\mathfrak{g}$. For Lie algebra-valued forms, it is convenient to define a bracket $[\cdot,\cdot]$, which combines the structure constants $f_{ab}{}^{c}$ of $\mathfrak{g}$ and the wedge product $\wedge$ of forms:
\begin{equation}\label{brack}
    [\omega,\eta]=\omega^{a}\wedge \eta^{b}\, f_{ab}{}^{c}\,T_{c}\; .
\end{equation}
The bracket inherits the graded symmetry of $\wedge$ and the antisymmetry of $f_{ab}{}^{c}$, and hence is graded antisymmetric with respect to the form degree:
\begin{equation}\label{wedgebracketsym}
    [\omega_p,\eta_q]=(-1)^{pq+1}[\eta_q,\omega_p]\; ,
\end{equation}
for a $\mathfrak{g}-$valued $p-$form $\omega_p$ and $q-$form $\eta_q$. For instance, for the typical case of two gauge fields one has the component expression
\begin{equation}
[A,A]=\big(f_{ab}{}^cA_\mu^aA_\nu^b\big)\,dx^\mu\wedge dx^\nu\,T_c\;,    
\end{equation}
which is symmetric in the two fields.
The de Rham differential $d=dx^\mu\del_\mu$ obeys a graded Leibniz rule with respect to $[\cdot,\cdot]$, i.e
\begin{equation}\label{leibbrad}
    d\,[\omega_p,\eta_q]=[d\omega_p,\eta_q]+(-1)^{p}[\omega_p,d\eta_q]\;,
\end{equation}
and the bracket $[\cdot,\cdot]$ obeys the graded Jacobi identity:
\begin{equation}\label{Jacgrad}
\big[[\omega_{p_1},\omega_{p_2}],\omega_{p_3}\big]+(-1)^{p_1(p_2+p_3)}\big[[\omega_{p_2},\omega_{p_3}],\omega_{p_1}\big]+(-1)^{p_3(p_1+p_2)}\big[[\omega_{p_3},\omega_{p_1}],\omega_{p_2}\big]=0 \;, 
\end{equation}
since $\mathfrak{g}$ is a Lie algebra and the wedge product is associative. 

The dynamics of the theory is encoded in the self-duality relation 
\begin{equation}\label{SDEQ}
(1-\star)F\equiv2\, P_{-}F=0\;,\end{equation}
written in terms of the anti self-dual projector, and we use conventions where the field strength $F$ of the gauge field $A$ is given by
\begin{equation}
    F=dA+\tfrac{1}{2}[A,A]\; .
\end{equation}
The theory is invariant under gauge transformations with a $\mathfrak{g}-$valued parameter $\Lambda=\Lambda^a\,T_a$:
\begin{equation}\label{gaugeper}
    \delta_{\Lambda} A=d\Lambda+[A,\Lambda]\;,
\end{equation}
since the field equation \eqref{SDEQ} is covariant:
\begin{equation}\label{eqn_gaugevar}
 \delta_\Lambda\big(P_-F\big)=[P_-F,\Lambda].   
\end{equation}
These transformations close on the gauge algebra bracket $[\Lambda_1,\Lambda_2]$, in that they obey 
\begin{equation}\label{closure}
[\delta_{\Lambda_{1}},\delta_{\Lambda_{2}}]A=\delta_{-[\Lambda_{1},\Lambda_{2}]}A\; ,
\end{equation}
where the square bracket on the left-hand side is the commutator of gauge variations and the one on the right-hand side is the bracket defined in \eqref{brack}. Equation \eqref{closure}  determines the gauge algebra of the theory, which is the Lie algebra of $\mathfrak{g}-$valued functions.

In the following we will show how all these familiar concepts from classical gauge theory are encoded in the data of an $L_\infty$ algebra.

\subsection{$L_\infty$ description of self-dual Yang-Mills}

$L_{\infty}$ algebras are mathematical structures underlying perturbative field theories, in that they encode their gauge structure, together with dynamics and consistent interactions \cite{Zwiebach:1992ie,Lada:1992wc,Hohm:2017pnh}. The reader familiar with the Batalin-Vilkovisky (BV) formulation of gauge theories will see the strong resemblance between the two descriptions which are, in fact, dual to one another \cite{Jurco:2018sby}.

An $L_{\infty}$ algebra is a $\mathbb{Z}-$graded vector space $\mathcal{X}=\bigoplus_{i}X_{i}$ equipped with a (possibly infinite) set of graded symmetric multilinear maps $B_{n}:\mathcal{X}^{\otimes n}\to \mathcal{X}$ of degree $|B_{n}|=1$, obeying a (possibly infinite) set of  quadratic relations (generalised Jacobi identities). Graded symmetry of the maps means that, given two adjacent inputs $\psi_i,\psi_j$, $B_{n}$ obeys
\begin{equation}\label{gradedsym}
    B_{n}(...,\psi_{i},\psi_{j},...)=(-1)^{\psi_{i}\psi_{j}}B_{n}(...,\psi_{j},\psi_{i},...)\;,
\end{equation}
where exponents in phase factors  always denote the degree of the corresponding element.
Denoting the unary bracket by $B_1\equiv\d$, the first $L_{\infty}$ relations read
\begin{equation}\label{linftyrels1}
\begin{split}
\d\big(\d(\psi)\big)&=0\;,\\[2mm]    
\d B_{2}(\psi_{1},\psi_{2})+B_{2}\big(\d(\psi_{1}),\psi_{2}\big)+(-1)^{\psi_{1}}B_{2}\big(\psi_{1},\d(\psi_{2})\big)&=0\;,\\[2mm]
B_{2}\big(B_{2}(\psi_{1},\psi_{2}),\psi_{3}\big)+(-1)^{\psi_{3}(\psi_1+\psi_2)}B_{2}\big(B_{2}(\psi_{3},\psi_{1}),\psi_{2}\big)+(-1)^{\psi_{1}(\psi_{2}+\psi_{3})}B_{2}\big(B_{2}(\psi_{2},\psi_{3}),\psi_{1}\big)&\\
+\d B_{3}(\psi_{1},\psi_{2},\psi_{3})+B_{3}(\d(\psi_{1}),\psi_{2},\psi_{3})+(-1)^{\psi_{1}}B_{3}(\psi_1,\d(\psi_{2}),\psi_{3})+(-1)^{\psi_{1}+\psi_{2}}B_{3}(\psi_{1},\psi_{2},\d(\psi_{3}))&=0
\end{split}    
\end{equation}
The first two relations in \eqref{linftyrels1} state that $\d$ is a differential, which acts as a derivation on the two-bracket $B_2$ (Leibniz rule). The last relation then states that $B_2$ obeys the Jacobi identity up to homotopy. This means that the failure of the two-bracket $B_{2}$ to be a (graded) Lie bracket is governed by the differential and three-bracket $B_3$ as above. In principle, there can be an infinite tower of similar higher relations in terms of higher $B_n$, which will not be needed in the following.

The first step to formulating gauge theories in the language of $L_{\infty}$ algebras is to assign a vector space to each object of the theory. For self-dual Yang-Mills, we have the graded vector space $\mathcal{X}^{\rm SDYM}=\bigoplus_{i=-1}^{1}X_{i}$ which consists of a space $X_{-1}$ of gauge parameters $\Lambda$, a space $X_{0}$ of gauge fields $A$, and a space $X_{1}$ of field equations\footnote{This is the space where field equations, and also sources, take value in an off-shell theory.} $E$. The $L_\infty$ degree of an element $\psi\in X_i$ is $|\psi|=i$. With these vector spaces, $\cX^{\rm SDYM}$ has the structure of a chain complex as follows:
\begin{equation}\label{SDYMXcomplex}
\begin{tikzcd}[row sep=2mm]
&X_{-1}\arrow{r}{\d} & X_{0}\arrow{r}{\d} & X_{1} \\
&\Lambda& A & E
\end{tikzcd}    \;,
\end{equation}
with $\d$ the differential of the theory, to be determined explicitly in the following. The gauge parameters are $\mathfrak{g}$-valued zero-forms, the fields are $\mathfrak{g}$-valued one-forms and the field equations are $\mathfrak{g}$-valued anti-self-dual two-forms.

In order to determine the nontrivial maps $B_{n}$ that act on $\mathcal{X}^{\rm SDYM}$, we expand the field equations \eqref{SDEQ} in powers of $A$: 
\begin{equation}\label{eomper}
    2\, P_{-}dA+P_{-}[A,A]=0\; .
\end{equation}
In $L_\infty$ language, the field equations take the form of a generalised Maurer-Cartan equation. Since the SDYM equations \eqref{eomper} are at most quadratic in $A$, this reduces to 
\begin{equation}\label{eoml}
\d A+\tfrac{1}{2}B_{2}(A,A)=0\;,
\end{equation}
with gauge transformations \eqref{gaugeper} given by
\begin{equation}\label{gaugel}
    \delta_\Lambda A=\d\Lambda+B_{2}(A,\Lambda)\; .
\end{equation}
Hence, one can identify the differential $\d$ with the gauge and kinetic operators defining the free theory, while the maps $B_{2}$ determine the interactions, nonlinear gauge symmetries and so on. In general, one can have higher brackets: Yang-Mills theory has cubic terms in the field equations, and hence a non-vanishing $B_{3}(A,A,A)$. Perturbative gravity would have an infinite number of brackets $B_n(h,\cdots,h)$ encoding the infinite interaction vertices of the graviton $h_{\mu\nu}$.

From the field equation \eqref{eomper} and the gauge transformation \eqref{gaugeper} one can completely determine the differential $\d$, together with the brackets $B_2(A,A)$ and $B_2(A,\Lambda)$:
\begin{align}\label{QB2}
\d(A)&=2\, P_{-}dA\, \in X_{1}\; ,& \d(\Lambda)&=d\Lambda\, \in X_{0}\nonumber\\
B_{2}(A_{1},A_{2})&=2\, P_{-}[A_{1},A_{2}]\, \in X_{1}\;, & B_{2}(A,\Lambda)&=[A,\Lambda]\, \in X_{0}\; .    
\end{align}
Notice that the differential and two-bracket act differently on elements of different vector spaces. In addition to the above, the remaining non vanishing brackets are between two gauge parameters and between a gauge parameter and a field equation. These can be determined from the closure of the gauge transformations, \eqref{closure}, and the gauge covariance of the field equations, \eqref{eqn_gaugevar}, respectively:
\begin{equation}\label{otherB2s}
    B_{2}(\Lambda_{1},\Lambda_{2})=-[\Lambda_{1},\Lambda_{2}]\, \in X_{-1}\; ,\quad B_{2}(\Lambda,E)=-[\Lambda,E]\, \in X_{1}\; .
\end{equation}
Let us briefly comment on the symmetry property of the brackets $B_2$: in the conventions we are using, they are graded symmetric, i.e. $B_2(\psi_1,\psi_2)=(-1)^{\psi_1\psi_2}B_2(\psi_2,\psi_1)$, with respect to the degree \eqref{SDYMXcomplex} in $\cX^{\rm SDYM}$. This should \emph{not} be confused with the symmetry \eqref{wedgebracketsym} of the bracket $[\cdot,\cdot]$. For instance, since $|A|=0$ and $|\Lambda|=-1$, one has $B_2(\Lambda,A)\equiv B_2(A,\Lambda)=[A,\Lambda]$, despite the fact that $[A,\Lambda]=-[\Lambda,A]$.

Having given all the maps acting on $\mathcal{X}^{\rm SDYM}$, we now turn to the $L_{\infty}$ relations \eqref{linftyrels1}, which encode the consistency of the theory.
In order to give a more concrete feeling of how the quadratic relations \eqref{linftyrels1} are equivalent to the usual concepts of gauge covariance, let us consider the field equation, written in the form \eqref{eoml}, and take a gauge variation under \eqref{gaugel}:
\begin{equation}\label{Loocov}
\begin{split}
\delta_\Lambda\Big(\d A+\tfrac12\,B_2(A,A)\Big)&=\d(\delta_\Lambda A)+\,B_2(\delta_\Lambda A,A)\\
&=\d\big(\d\Lambda+B_2(A,\Lambda)\big)+B_2\big(\d\Lambda+B_2(A,\Lambda),A\big)\\
&=\d^2(\Lambda)+\d B_2(A,\Lambda)+B_2\big(\d\Lambda,A\big)+B_2\big(B_2(\Lambda,A),A\big)\;.
\end{split}    
\end{equation}
For the equation of motion to be gauge covariant, the right-hand side should be proportional to the original field equation $\d A+\frac12\,B_2(A,A)$. To this end, we rewrite it as
\begin{equation}
\begin{split}
\delta_\Lambda\Big(\d A+\tfrac12\,B_2(A,A)\Big)=&-B_2\Big(\d A+\tfrac12\,B_2(A,A),\Lambda\Big)\\
&+\d^2(\Lambda)\\
&+\d B_2(A,\Lambda)+B_2\big(\d A,\Lambda\big)+B_2\big(A,\d\Lambda\big)\\
&+\frac12\Big[B_2\big(B_2(A,A),\Lambda\big)+2\,B_2\big(B_2(\Lambda,A),A\big)\Big] \;,  
\end{split}    
\end{equation}
where we split the last three lines in powers of $A$. One can see that demanding gauge covariance of the field equations, i.e. demanding that the last three lines above vanish separately order by order in $A$, is the same as imposing the quadratic relations \eqref{linftyrels1} for the case of inputs $(\psi_1,\cdots,\psi_n)$ given by one parameter $\Lambda$ and $n-1$ fields $A$. Similarly, the quadratic relations for other assignments of inputs (unless trivial by degree), encode the closure of the gauge algebra and consistency of the latter with covariance of the equations.
All the nontrivial relations \eqref{linftyrels1} can be easily checked from the explicit forms \eqref{QB2} and \eqref{otherB2s} of the differential and two-brackets, upon using the Leibniz \eqref{leibbrad} and graded Jacobi \eqref{Jacgrad} identities of $d$ and $[\cdot,\cdot]$, together with $[\star\omega_p,\Lambda]=\star[\omega_p,\Lambda]$ which is valid for a $\mathfrak{g}-$valued zero-form $\Lambda$.

Notice that the $L_\infty$ relations \eqref{linftyrels1} in this case are obeyed with no higher brackets: $B_n=0$ for $n>2$. The $L_\infty$ algebra $\cX^{\rm SDYM}$ associated to self-dual Yang-Mills is thus strict, i.e. it is a differential graded Lie algebra. In the next section we will strip off the color algebra $\mathfrak{g}$ from $\cX^{\rm SDYM}$, and build the kinematic algebra on the remaining space.

\section{Gauge independent kinematic algebra and double copy}\label{Sec:BVoo}

In this section we take the differential graded Lie algebra (strict $L_\infty$) of self-dual Yang-Mills and disentangle its color degrees of freedom from the kinematic ones, in order to construct a kinematic algebra on the color-stripped space. We then implement the gauge invariant double copy prescription of \cite{Bonezzi:2022bse} and recover self-dual gravity at the free level. In the next section we will show how to construct the full theory by applying the same procedure in light-cone gauge.

\subsection{Color-stripping}
\label{sub:Color-Stripping}

In the graded vector space $\cX^{\rm SDYM}$ associated to self-dual Yang-Mills theory all elements, such as gauge parameters, fields etc., take value in the color Lie algebra $\mathfrak g$.
The $L_{\infty}$ algebra $\cX^{\rm SDYM}$ thus takes the form of a tensor product:
\begin{equation}
    \mathcal{X}^{\rm{SDYM}}=\mathcal{K}\otimes \mathfrak{g}\;,
\end{equation}
where the kinematic space $\mathcal{K}$ is an infinite dimensional graded vector space that contains all the spacetime dependence and kinematic information of the theory. One can expand an arbitrary element $\psi(x)$ of $\mathcal{X}^{\rm{SDYM}}$ in a basis $\{T_a\}$ of $\mathfrak g$, and write it as
\begin{equation}\label{elementfact}
    \psi(x)=u^{a}(x)\otimes T_{a}\; ,\;\, u^{a}(x)\in\mathcal{K}\; ,\;\, T_{a}\in \mathfrak{g}\; .
\end{equation}
The $u^{a}(x)$ should be thought of as \textit{color-stripped} objects only carrying kinematic information. For instance, when considering scattering amplitudes, $\psi$ is typically a gauge field $A$ representing an external particle and its associated $u$ carries the polarization vector and momentum of the gluon, e.g. ${u=\epsilon_\mu(p)\,e^{ip\cdot x}dx^\mu}$. Notice that we dropped the color index $a$. Indeed, in this framework, the color information is encoded in the relation between the $L_\infty$ bracket $B_2$ and the kinematic product $m_2$ to be defined below. 

The differential $\d$ does not act on the color degrees of freedom and can thus be defined directly as an operator acting on $\cK$ by
\begin{equation}
\d\big(\psi(x)\big)=\d\big(u^{a}(x)\big)\otimes T_{a}\; ,
\end{equation}
while the bracket $B_{2}$ can be factorised as\footnote{The phase prefactor is chosen so that $m_2$ is graded symmetric with respect to the degree of $\cK$.} \begin{equation}
    B_{2}(\psi_{1},\psi_{2})=(-1)^{\psi_{1}}m_{2}(u_{1}^{a},u_{2}^{b})\otimes f_{ab}{}^{c}\,T_{c}\; ,
\end{equation}
where $m_{2}$ is a kinematic product. This factorization is obvious from the definition of $[\cdot,\cdot]$ in terms of the wedge product of forms and the structure constants in equation \eqref{brack}.
$\mathcal{K}$ is a graded vector space and in our conventions its elements have \textit{kinematic degree}
\begin{equation}
|u|_{\mathcal{K}}=|\psi|_{\mathcal{X}^{\rm{SDYM}}}+1\; ,
\end{equation}
while the kinematic maps $\d$ and $m_{2}$ have degree one and zero, respectively. The kinematic degree coincides with the form degree and makes $m_2$ graded symmetric with respect to it: $m_2(u_1,u_2)=(-1)^{u_1u_2}m_2(u_2,u_1)$. More precisely, $\mathcal{K}$ is given by the direct sum
\begin{equation}
    \mathcal{K}=\bigoplus_{i=0}^{2}K_{i}\; ,
\end{equation}
where $K_{0}$ is the space of zero-forms on $\mathbb{R}^{4}$, $K_{1}$ the space of one-forms, and $K_{2}$ is the space of anti-self-dual two-forms. Equipped with the differential $\d$, one then has a de Rham-like chain complex
\begin{equation}\label{Kdiagram}
\begin{tikzcd}[row sep=2mm]
K_{0}\arrow{r}{\d} & K_{1}\arrow{r}{\d} & K_{2} \\
 \lambda& \cA & \cE
\end{tikzcd}    \;,
\end{equation}
where we changed the font of the elements to emphasise that now we are dealing with color stripped quantities. With an abuse of terminology, we still refer to $\lambda$ as gauge parameters, to $\cA$ as fields etc. The explicit action of the differential and the two-product on the elements of $\mathcal{K}$ is 

\begin{align}\label{m2def}
    \d\lambda&=d\lambda \in K_{1}\;, &\; \; \d\cA&=2\, P_{-}d\cA\in K_{2}\nonumber\\
    m_{2}(\lambda_{1},\lambda_{2})&=\lambda_{1}\wedge \lambda_{2}\in K_{0}\; , &\;\; m_{2}(\lambda,\cA)&=\lambda\wedge \cA\in K_{1}\\
    m_{2}(\cA_{1},\cA_{2})&=2\, P_{-}\big(\cA_{1}\wedge\cA_{2}\big)\in K_{2}\; , &\;\; m_{2}(\lambda,\cE)&=\lambda\wedge\cE\in K_{2}\;\nonumber ,
\end{align}
where we remind the reader that $2P_-=1-\star$.
The vector space $\mathcal{K}$, endowed with the differential $\d$ and the two-product $m_{2}$, forms a differential graded commutative algebra. In these algebras, the differential and the two-product obey the following relations:
\begin{equation}\label{cinfty}
\begin{split}
    \d(\d(u))=0\;,\\
    \d m_{2}(u_{1},u_{2})-m_{2}(\d u_{1},u_{2})-(-1)^{u_{1}}m_{2}(u_{1},\d u_{2})=0\; ,\\
    m_{2}(m_{2}(u_{1},u_{2}),u_{3})-m_{2}(u_{1},m_{2}(u_{2},u_{3}))=0\; .
\end{split}
\end{equation}
The first line is, again, the nilpotency of the differential, the second line is the Leibniz rule of $\d$ with respect to $m_{2}$, and the last line is the associativity of the two-product. The relations above can be checked straightforwardly by using the properties of the de Rham differential, of the the wedge product and of the Hodge star operator. Let us stress that, at this stage, the algebraic structure encoded in \eqref{cinfty} is an \emph{associative} algebra, which does \emph{not} have a Lie bracket.

\subsection{Construction of the kinematic algebra}

It has been shown that theories like Chern-Simons \cite{Ben-Shahar:2021zww,Borsten:2022vtg} and pure Yang-Mills \cite{Reiterer:2019dys,Bonezzi:2022bse} have underlying kinematic algebras that are generalizations of  Batalin-Vilkovisky algebras, named BV$_{\infty}^{\square}$ algebras, which were first constructed in \cite{Reiterer:2019dys}. The kinematic BV$_{\infty}^{\square}$ algebra of Yang-Mills is, at least up to quartic interactions, the structure responsible for the consistency of a double copy prescription that yields gravity in the form of double field theory \cite{Bonezzi:2022bse}. Hence, it is naturally expected that self-dual Yang-Mills also possesses a kinematic algebra of this type. This is an off-shell and gauge independent generalization of the algebra of area-preserving diffeomorphisms discovered by Monteiro and O'Connell \cite{Monteiro:2011pc} in light-cone gauge, as we will show in the following.

To this end, consider the differential graded commutative algebra constructed in the previous section, defined as the space $\mathcal{K}$ endowed with the differential $\d$ and the two-product $m_{2}$ acting as in \eqref{m2def} and obeying the relations in \eqref{cinfty}.
The crucial step to unveiling a richer algebraic structure is to equip $\cK$ with an additional differential $b$. This has opposite degree to $\d$, namely $|b|=-1$ and acts on the chain complex \eqref{Kdiagram} as
\begin{equation}
\begin{tikzcd}[row sep=4mm]
 K_{0}\arrow[r, "\d "]&\arrow[l, bend left=50, "b"] K_{1}\arrow[r, "\d "]&\arrow[l, bend left=50, "b"]  K_{2} \\
\lambda& \cA & \cE&\;.
\end{tikzcd}    
\end{equation}
We require $b$ to be nilpotent:
$b^{2}=0$, and to obey the defining relation
\begin{equation}\label{commbox}
\d b+b\,\d=\B\;,
\end{equation}
where we denote the Laplacian as $\B=\del^\mu\del_\mu$.
The reason for considering such an operator can be traced back to scattering amplitudes: when computing amplitudes, one builds Feynman diagrams out of vertices and propagators of a gauge fixed theory. Having an operator $b$ obeying the above relations, one can gauge fix the theory by imposing $b\cA=0$ on fields and, thanks to \eqref{commbox}, write a propagator as $\frac{b}{\B}$ acting on the space of equations (or sources) $\cE$.  

Such an operator has been constructed for different gauge theories, like Chern-Simons and Yang-Mills \cite{Reiterer:2019dys,Ben-Shahar:2021zww,Borsten:2022vtg,Bonezzi:2022bse}.  We choose the $b$ operator in this case to be the adjoint of the de Rham differential:
\begin{equation}
\label{b_define}
b=d^{\dagger}:=-\star d\star\; ,
\end{equation}
so that $(d^\dagger\omega)_{\mu_1\cdots\mu_{p-1}}=\del^\nu\omega_{\nu\mu_1\cdots\mu_{p-1}}$ in components and $d\,d^\dagger+d^\dagger\,d=\B$.
The operator $b$ has degree minus one as it decreases the form degree by one and it is indeed nilpotent.
The defining relation \eqref{commbox} is not obvious because $\d $ is, in general, not just the de Rham differential. As an example, let us check that \eqref{commbox} is obeyed on elements of the space $K_{1}$, namely
\begin{equation}
\begin{split}
    \d \big(b(\cA)\big)+b\big(\d (\cA)\big)&=dd^{\dagger}\cA+2\, d^{\dagger}P_{-}d\cA\\
    &=dd^{\dagger}\cA+d^{\dagger}d\cA\\
    &=\Box \cA\; ,
\end{split}
\end{equation}
where in the second line we used $d^{\dagger}\star d=0$. A similar computation for $\lambda$ and $\cE$ then proves that \eqref{commbox} is obeyed for all elements of $\mathcal{K}$. 

Having this new differential at our disposal, one can study its compatibility with the associative product $m_2$.
If $b$ does not obey the Leibniz rule with respect to $m_{2}$, its failure to do so can be used to define a graded symmetric bracket $b_{2}$ as 
\begin{equation}\label{b2def}
b_{2}(u_{1},u_{2}):=b\,m_{2}(u_{1},u_{2})-m_{2}(bu_{1},u_{2})-(-1)^{u_{1}}m_{2}(u_{1},bu_{2})\; ,
\end{equation}
with intrinsic degree $|b_2|=-1$. In an amplitudes context, the bracket $b_2(\cA_1,\cA_2)$ between color-stripped fields gives the contribution to the kinematic numerator arising from a cubic vertex joining the external particles $1$ and $2$. 
In physics, the typical example where one has both a bracket and a product is the Poisson algebra of Hamiltonian mechanics. There, the Poisson bracket $\{\cdot,\cdot\}$ obeys the Leibniz rule with respect to the pointwise product of functions:
\begin{equation}
\big\{fg,h\big\}=\big\{g,h\big\}f+\big\{f,h\big\}g\;,    
\end{equation}
which is the (Poisson) compatibility between the two. Similarly, in the graded case the compatibility between the bracket $b_2$ and the product $m_2$ would read 
\begin{equation}\label{compa}
    b_{2}(m_{2}(u_{1},u_{2}),u_{3})=(-1)^{u_{1}(u_{2}+u_{3})}m_{2}(b_{2}(u_{2},u_{3}),u_{1})
    +(-1)^{u_{3}(u_{1}+u_{2})}m_{2}(b_{2}(u_{3},u_{1}),u_{2})\;.
\end{equation}
An algebra which satisfies the above is called a \textit{strict} BV algebra. An important difference, compared to the standard Poisson algebra of Hamiltonian mechanics, is that the kinematic bracket $b_2$ is \emph{not} an independent object, since it is derived from $b$ and $m_2$. In particular, this implies that \emph{if} the Poisson compatibility \eqref{compa} holds, $b_2$ is guaranteed to obey the (graded) Jacobi identity, thus making it into a graded Lie bracket. This is proven by acting with $b$ on the relation \eqref{compa}, which turns all the products $m_2$ into brackets $b_2$ and converts \eqref{compa} into the Jacobi identity of $b_2$ (the details of the proof can be found in \cite{Bonezzi:2022bse}): 
\begin{equation}
\begin{split}
    b_{2}(b_{2}(u_{1},u_{2}),u_{3})&+(-1)^{u_{1}(u_{2}+u_{3})}b_{2}(b_{2}(u_{2},u_{3}),u_{1})\\
    &+(-1)^{u_{3}(u_{1}+u_{2})}b_{2}(b_{2}(u_{3},u_{1}),u_{2})=0\;.
\end{split}
\end{equation}
In Chern-Simons theory, the compatibility condition \eqref{compa} does hold, and the corresponding $b_2$ is a genuine graded Lie bracket \cite{Ben-Shahar:2021zww,Borsten:2022vtg,Bonezzi:2022bse}. Geometrically, this kinematic bracket generates the Schouten-Nijenhuis algebra of polyvector fields in three dimensions, which contains three dimensional diffeomorphisms as a subalgebra. It is important to stress that even in the case where $b_2$ is a Lie bracket, the kinematic algebra is strictly larger than a Lie algebra. Indeed, in Chern-Simons the triplet $(b,m_2,b_2)$ form a BV algebra, with $b_2$ generating the Lie subsector.

So far, we have not discussed the property of the kinematic bracket $b_2$ with respect to the original differential $\d$. Given the defining property $\d b+b\d =\square$, it follows that $\d$
obeys the Leibniz rule with respect to $b_{2}$ only \textit{modulo box}, meaning
\begin{equation}\label{DefLeibniz}
\begin{split}
    \d b_{2}(u_{1},u_{2})-b_{2}(\d u_{1},u_{2})-(-1)^{u_{1}}b_{2}(u_{1},\d u_{2})=\Box m_{2}(u_{1},u_{2})&-m_{2}(\Box u_{1},u_{2})\\
    &-m_{2}(u_{1},\Box u_{2})\; .
\end{split}
\end{equation}
This is why, upon including $\d$, the kinematic algebra of Chern-Simons theory was termed a ${\rm BV}^\B$ algebra in \cite{Borsten:2022vtg}.

Having discussed the simpler strict case, where \eqref{compa} holds, we now turn to the case of self-dual Yang-Mills. As we will show in the following, the compatibility relation \eqref{compa} does not hold strictly, but up to homotopy, meaning 
\begin{equation}\label{hompoi}
\begin{split}
    b_{2}(m_{2}(u_{1},u_{2}),u_{3})&-(-1)^{u_{1}(u_{2}+u_{3})}m_{2}(b_{2}(u_{2},u_{3}),u_{1})\\
    &-(-1)^{u_{3}(u_{1}+u_{2})}m_{2}(b_{2}(u_{3},u_{1}),u_{2})=[\d ,\theta_{3}](u_{1},u_{2},u_{3})\;,
\end{split}
\end{equation}
where we introduced the graded commutator $[\d ,\theta_{3}]$, defined as 
\begin{equation}\label{commutator}
\begin{split}
    [\d ,\theta_{3}](u_{1},u_{2},u_{3}):=\d \theta_{3}(u_{1},u_{2},u_{3})-\theta_{3}(\d u_{1},u_{2},u_{3})&-(-1)^{u_{1}}\theta_{3}(u_{1},\d u_{2},u_{3})\\
    &-(-1)^{u_{1}+u_{2}}\theta_{3}(u_{1},u_{2},\d u_{3})\; .
\end{split}
\end{equation}
The $\theta_3$ above is a trilinear map playing the role of a homotopy for the Poisson identity, similarly to the case of $L_\infty$ algebras in \eqref{linftyrels1}, where the bracket $B_3$ is a homotopy for the Jacobi identity.
Algebras of this type were named BV$_{\infty}^{\square}$ algebras in \cite{Reiterer:2019dys} and can potentially have an infinite number of higher maps, e.g $\theta_{4}$, and thus higher relations. In this paper we will restrict our construction to trilinear maps and hence this is the only relation that we need for our purposes. Indeed,
as we have discussed above, the Poisson relation \eqref{hompoi} is sufficient to determine the identity obeyed by $b_2$. Since the compatibility \eqref{hompoi} is not strict, $b_2$ does not obey a strict Jacobi identity, but rather \cite{Bonezzi:2022bse}
\begin{equation}
\begin{split}
  b_{2}(b_{2}(u_{1},u_{2}),u_{3})&+(-1)^{u_{1}(u_{2}+u_{3})}b_{2}(b_{2}(u_{2},u_{3}),u_{1})+(-1)^{u_{3}(u_{1}+u_{2})}b_{2}(b_{2}(u_{3},u_{1}),u_{2})\\
  &+[\d ,b_{3}](u_{1},u_{2},u_{3})+[\B,\theta_3](u_1,u_2,u_3)=0\;,    
\end{split}    
\end{equation}
with a three-bracket $b_3$ which is also derived: $b_3=-[b,\theta_3]$, and a $\B-$deformation controlled by $\theta_3$. Here we are using a compact commutator notation similar to \eqref{commutator}:
\begin{equation}
\begin{split}
[\d ,b_{3}](u_{1},u_{2},u_{3}):=\; &\d b_{3}(u_{1},u_{2},u_{3})+b_{3}(\d u_{1},u_{2},u_{3})+(-1)^{u_{1}}b_{3}(u_{1},\d u_{2},u_{3})\\
&+(-1)^{u_{1}+u_{2}}b_{3}(u_{1},u_{2},\d u_{3})\;,\\
[\B ,\theta_{3}](u_{1},u_{2},u_{3}):=\; &\B \theta_{3}(u_{1},u_{2},u_{3})-\theta_{3}(\B u_{1},u_{2},u_{3})-\theta_{3}(u_{1},\B u_{2},u_{3})-\theta_{3}(u_{1},u_{2},\B u_{3})\;,\\
[b,\theta_{3}](u_{1},u_{2},u_{3}):=\; &b\, \theta_{3}(u_{1},u_{2},u_{3})-\theta_{3}(b u_{1},u_{2},u_{3})-(-1)^{u_{1}}\theta_{3}(u_{1},b u_{2},u_{3})\\
    &-(-1)^{u_{1}+u_{2}}\theta_{3}(u_{1},u_{2},b u_{3})\; .
\end{split}    
\end{equation}
One can thus see that, similarly to the full Yang-Mills case, the kinematic algebra of off-shell and gauge invariant SDYM theory  does not contain a Lie algebra as a consistent subsector, in contrast to Chern-Simons.

We now turn to showing that, indeed, to trilinear order there is a BV$_{\infty}^{\square}$ kinematic algebra underlying self-dual Yang-Mills theory. 
The derived two-bracket $b_{2}$ can be computed using \eqref{b2def} and the explicit form of the bracket acting on the different elements of $\mathcal{K}$ reads
\begin{align}\label{b2}
    b_{2}(\lambda,\cA)&=d^{\dagger}(\cA\wedge\lambda)-d^{\dagger}A\wedge\lambda\in K_{0}\; ,&\;\; b_{2}(\cA_{1},\cA_{2})&=2\, d^{\dagger} P_{-}(\cA_{1}\wedge \cA_{2})-2\, d^{\dagger}\cA_{[1}\wedge \cA_{2]}\in K_{1}\; ,\nonumber\\
    b_{2}(\lambda,\cE)&=\star(\cE\wedge d\lambda)\in K_{2}\; ,&\;\; b_{2}(\cA,\cE)&=2\, P_{-}(\cA\wedge d^{\dagger}\cE)-d^{\dagger}\cA\wedge \cE\in K_{2}\; ,
\end{align}
with all other brackets vanishing.
Given the differential $\d$ and two-product $m_2$ in \eqref{m2def} and the two-bracket $b_2$ above, one has to prove the
compatibility relation \eqref{hompoi}. To that end, we need to find the explicit expression of the homotopy map $\theta_{3}$ acting on the different elements of $\mathcal{K}$. In order to find $\theta_{3}$, it suffices to compute the left hand side of \eqref{hompoi}. Then, one should be able to identify the differential acting on various elements and determine $\theta_{3}$. As an example, let us consider two color stripped gauge fields and one gauge parameter:
\begin{equation}
    b_{2}(m_{2}(\cA_{1},\cA_{2}),\lambda)-2\, m_{2}(b_{2}(\lambda,\cA_{[1}),\cA_{2]})\stackrel{!}{=}[\d ,\theta_{3}](\cA_{1},\cA_{2},\lambda)\; .
\end{equation}
Writing the left hand side in terms of the maps shown in equations \eqref{m2def} and \eqref{b2} yields
\begin{equation}\label{twoterm}
    2\, \star \Big\{ P_{-} \big( \cA_{1}\wedge \cA_{2}\big)\wedge d\lambda \Big\}-2\, \Big\{ d^{\dagger}\big( \cA_{[1}\wedge \lambda \big)\wedge \cA_{2]}-d^{\dagger}\cA_{[1}\wedge \cA_{2]}\wedge \lambda \Big\}\stackrel{!}{=}[\d ,\theta_{3}](\cA_{1},\cA_{2},\lambda)\;.
\end{equation}
The first term in the above equation can be rewritten as
\begin{equation}
\begin{split}
 2\, \star \Big\{ P_{-} \big( \cA_{1}\wedge \cA_{2}\big)\wedge d\lambda \Big\}&=\star\Big\{ \cA_{1}\wedge \cA_{2}\wedge d\lambda \Big\}-\star d\Big\{ \star\big(\cA_{1}\wedge \cA_{2}\big)\wedge \lambda \Big\}\\
 &+\star\Big\{ d\star\big( \cA_{1}\wedge \cA_{2} \big)\wedge \lambda \Big\}\\
 &=\star\Big\{ \cA_{1}\wedge \cA_{2}\wedge d\lambda \Big\}+d^{\dagger}\Big\{ \cA_{1}\wedge \cA_{2}\wedge \lambda \Big\}\\
 &-d^{\dagger}\big(\cA_{1}\wedge \cA_{2}\big)\wedge \lambda\; ,
\end{split}
\end{equation}
where we used the fact that for a zero-form $\lambda$ and a $p$-form $\omega$ the following is obeyed: 
$\star(\omega\wedge\lambda)=\star\omega\wedge\lambda$. Using that $d^{\dagger}$ is a \textit{second order} operator with respect to the wedge product of $p$-forms, namely
\begin{equation}
\begin{split}
    d^{\dagger}(u\wedge v\wedge w)=&-d^{\dagger}u\wedge v\wedge w-(-1)^{u}u\wedge d^{\dagger}v\wedge w-(-1)^{u+v}u\wedge v\wedge d^{\dagger}w\\
    &+d^{\dagger}\big(u\wedge v\big)\wedge w+(-1)^{u}u\wedge d^{\dagger}\big( v\wedge w \big)+(-1)^{uv+v}v\wedge d^{\dagger}\big(u\wedge w \big)\;,
\end{split}
\end{equation}
the two terms on the left hand side of equation \eqref{twoterm} reduce to
\begin{equation}
    \star\Big\{ \cA_{1}\wedge \cA_{2}\wedge d\lambda \Big\}\stackrel{!}{=}[\d ,\theta_{3}](\cA_{1},\cA_{2},\lambda)\; .
\end{equation}
Notice that none of the terms on the left hand side of the above equation have the differential acting on the fields, nor on the entire expression, but one can identify the differential acting on $\lambda$ ($\d (\lambda)=d\lambda$). Thus, one can infer that both $\theta_{3}(\cA_{1},\cA_{2},\lambda)$ and $\theta_{3}(\cE,\cA,\lambda)$ vanish, while 
\begin{equation}
    \star\Big\{ \cA_{1}\wedge \cA_{2}\wedge d\lambda \Big\}\stackrel{!}{=}-\theta_{3}(\cA_{1},\cA_{2},\d \lambda)\; ,
\end{equation}
determines (recalling that $\d \lambda\in K_1$)
\begin{equation}
    \theta_{3}(\cA_{1},\cA_{2},\cA_{3})=-\star \big(\cA_{1}\wedge \cA_{2}\wedge \cA_{3}\big)\; .
\end{equation}
Following analogous computations for all degrees, one finds that the only non-trivial $\theta_{3}$ are
\begin{equation}\label{theta3}
\begin{split}
\theta_{3}(\cA_{1},\cA_{2},\cA_{3})&=-\star \big(\cA_{1}\wedge \cA_{2}\wedge \cA_{3}\big)\in K_{1}\; ,\\
\theta_{3}(\cE,\cA_{1},\cA_{2})&=2\, P_{-}\Big\{ \star\big(\cE\wedge\cA_{[1}\big)\wedge \cA_{2]}  \Big\}\in K_{2}\; ,
\end{split}
\end{equation}
and hence we conclude that the off-shell and gauge invariant kinematic algebra of self-dual Yang-Mills is a BV$_{\infty}^{\square}$ algebra to trilinear order. 

\subsection{Gauge invariant double copy}

In this section we turn to constructing linearised self-dual gravity using the double copy prescription of \cite{Bonezzi:2022bse}. 
Having identified the kinematic algebra of $\cK$ up to trilinear maps, one can construct the gauge invariant double copy up to cubic terms in the field equations. In order to illustrate how the gauge independent double copy prescription works, we restrict here to the free theory, where one can easily make contact with self-dual gravity.  

Our starting point is the $L_{\infty}$-algebra of SDYM written as the tensor product
\begin{equation}
\mathcal{X}^{\rm{SDYM}}=\cK\otimes \mathfrak{g}\; .
\end{equation}
The next step in this prescription is to replace the color Lie algebra $\mathfrak{g}$ by another copy of the kinematic algebra $\bar\cK$ with its own coordinates $\bar x^{\bar \mu}$. Doing so leads to the tensor product space
\begin{equation}
    \mathcal{X}^{\rm DC}=\cK\otimes \bar\cK\; .
\end{equation}
The elements of $\mathcal{X}^{\rm{DC}}$ are thus functions of the doubled coordinates $(x^{\mu},\bar x^{\bar\mu})$. This was originally proposed in \cite{Diaz-Jaramillo:2021wtl,Bonezzi:2022yuh}
to construct double field theory \cite{Hull:2009mi,Hohm:2010jy,Hohm:2010pp}, a T-duality covariant reformulation of $\cN=0$ supergravity\footnote{We commonly refer as such to the bosonic theory including the graviton, the $B-$field and dilaton, which is common to all the low-energy effective actions of closed strings.}, as the double copy of Yang-Mills.
As discussed in \cite{Diaz-Jaramillo:2021wtl,Bonezzi:2022yuh}, in order for the double copy to be consistent one has to impose a constraint on the functional dependence of fields in the doubled space. In double field theory terminology, here we impose the \textit{strong constraint} on the elements $\Psi_{i}(x,\bar x)\in \mathcal{X}^{\rm{DC}}$:
\begin{equation}\label{strong}
    \square \Psi =\bar{\square}\Psi\;,\;\; \del_{\mu}\Psi_{1}\del^{\mu}\Psi_{2}=\del_{\bar\mu}\Psi_{1}\del^{\bar \mu}\Psi_{2}\; ,
\end{equation}
where the derivatives with barred indices $\del_{\bar\mu}$ are with respect to $\bar x^{\bar \mu}$, and the barred Laplacian is given by $\bar\square\equiv \del^{\bar\mu}\del_{\bar\mu}$. The strong constraint implies that $\B=\bar\B$ acting on arbitrary products of functions, so that one can view $\B\equiv\bar\B$ as operators. Effectively, the strong constraint eliminates the dependence on half of the coordinates and its simplest solution is the so-called supergravity solution where one identifies the two sets of indices and coordinates, i.e $\mu\equiv \bar\mu$ and $x^{\mu}\equiv \bar x^{\bar\mu}$. For the moment, however, we will keep the dependence on both sets of coordinates, as well as the additional set of indices, as a bookkeeping device to keep track of the index factorization property of the double copy. 

The double copy space $\mathcal{X}^{\rm{DC}}$ has a grading inherited from the grading of the two copies of the kinematic vector spaces $\cK$ and $\bar \cK$. More precisely, $\mathcal{X}^{\rm{DC}}$ is the following direct sum:
\begin{equation}
    \mathcal{X}^{\rm DC}=\bigoplus_{i=0}^{4} X^{\rm DC}_{i}\;,
\end{equation}
and each one of the vector subspaces $X^{\rm{DC}}_{i}$ is expressed in terms of the single copy spaces $K_{i}$ and $\bar K_{i}$ as
\begin{equation}
\begin{split}
    X^{\rm DC}_{0}&=K_{0}\otimes \bar K_{0}\; ,\\
    X^{\rm DC}_{1}&=\big(K_{0}\otimes \bar K_{1}\big)\oplus \big(K_{1}\otimes \bar K_{0}\big)\; ,\\
    X^{\rm DC}_{2}&=\big(K_{0}\otimes \bar K_{2}\big)\oplus \big(K_{1}\otimes \bar K_{1}\big)\oplus \big(K_{2}\otimes \bar K_{0}\big)\; ,\\
    X^{\rm DC}_{3}&=\big(K_{1}\otimes \bar K_{2}\big)\oplus \big(K_{2}\otimes \bar K_{1}\big)\; ,\\
    X^{\rm DC}_{4}&=K_{2}\otimes \bar K_{2}\; .\\
\end{split}
\end{equation}
In order to encode a field theory in this framework, we need to define the $L_{\infty}$ brackets that determine the dynamics, interactions and gauge structure of the theory. In this section we restrict our attention to linear dynamics and hence we only need to construct a differential $B_{1}$ which encodes the linearised gravity theory that follows from our double copy prescription. As we prove in the following, the double copy leads to pure self-dual gravity. In view of this, we define the differential $B_{1}$ in terms of the differentials of the single copy theories as 
\begin{equation}
    B_{1}=\d\otimes 1+1\otimes\bar\d\; .
\end{equation}
This new differential is nilpotent. Indeed, given that both $\d$ and $\bar \d$ are nilpotent and odd, we have
\begin{equation}
    B_{1}^{2}=\d^{2}\otimes 1+\d\otimes\bar\d-\d\otimes\bar\d+1\otimes\bar\d^{2}=0\; ,
\end{equation}
where we used $(1\otimes\bar\d)(\d\otimes 1)=-\d\otimes\bar\d$, which follows from both operators having odd degree. At linear order, the only relevant $L_{\infty}$ relation is nilpotency of the differential, and hence the vector space $\mathcal{X}^{\rm{DC}}$ equipped with $B_{1}$ as defined above encodes a consistent free field theory. 

In order to have a better understanding of the physical significance of the double copy space, it will be useful to organize the theory in the following chain complex:
\begin{equation}
\begin{tikzcd}[row sep=2mm]
X^{\rm DC}_{0}\arrow{r}{B_{1}} & X^{\rm DC}_{1}\arrow{r}{B_{1}} & X^{\rm DC}_{2}\arrow{r}{B_{1}} & X^{\rm DC}_{3}\arrow{r}{B_{1}} & X^{\rm DC}_{4} \\
\chi & \Lambda & \Psi & \Sigma & \cN
\end{tikzcd}    \;.
\end{equation}
The elements of the different subspaces are $(p,q)$ bi-forms ($p-$forms in $dx$ and $q-$forms in $d\bar x$), which in components read
\begin{equation}\label{listofthings}
\begin{split}
\chi&=\chi(x,\bar x)\; ,\\
\Lambda&=\lambda+\bar\lambda=-\lambda_{\mu}(x,\bar x)\, dx^{\mu}+\bar\lambda_{\bar\mu}(x,\bar x)\, d\bar x^{\bar\mu}\; ,\\
\Psi&=f+e+\bar f=\tfrac{1}{2}\, f_{\mu\nu}(x,\bar x)\, dx^{\mu}\wedge dx^{\nu}+e_{\mu\bar\nu}(x,\bar x)\, dx^{\mu}\otimes d\bar x^{\bar\nu}+\tfrac{1}{2}\, \bar f_{\bar\mu\bar\nu}(x,\bar x)\, d\bar x^{\bar \mu}\wedge d\bar x^{\bar\nu}\; ,\\
\Sigma&=\omega+\bar \omega=\tfrac{1}{2}\, \omega_{\mu\nu\bar\rho}(x,\bar x)\, dx^{\mu}\wedge dx^{\nu}\otimes d\bar x^{\bar \rho}+\tfrac{1}{2}\, \bar\omega_{\mu\bar\nu\bar\rho}(x,\bar x)\, d x^{\mu}\otimes d \bar x^{\bar\nu}\wedge d\bar x^{\bar\rho}\; ,\\
\cN&=\tfrac{1}{4}\, N_{\mu\nu\bar\rho\bar\sigma}(x,\bar x)\, dx^{\mu}\wedge dx^{\nu}\otimes d\bar x^{\bar\rho}\wedge d\bar x^{\bar \sigma}\; .
\end{split}
\end{equation}
In the above decomposition, all $(2,q)$ and $(p,2)$ components are anti-self-dual in the two-form sector, e.g. $f=-\star f$ and $\bar f=-\bar\star\bar f$.
The space $X_{0}^{\rm{DC}}$ contains gauge-for-gauge parameters, the space $X^{\rm{DC}}_{1}$ is the space of gauge parameters and $X^{\rm{DC}}_{2}$ is the space of fields. The elements of $X_{3}^{\rm{DC}}$ are field equations and $X^{\rm{DC}}_{4}$ is a space of identities. Notice that the degree of the elements of $\mathcal{X}^{\rm{DC}}$ coincides with their total bi-form degree $p+q$. 

Let us turn to the action of the differential on objects with different degree. For instance, the fields $\Psi$ consist of a tensor fluctuation $e_{\mu\bar\nu}$ (containing the graviton and $B$-field off-shell), and anti-self-dual two-forms $f_{\mu\nu}$ and $\bar f_{\bar\mu\bar\nu}$.
The free field equations correspond to $B_{1}(\Psi)=0$,  which yields
\begin{equation}\label{firstoreoms}
\begin{split}
2\, P_{-}\, d e+\bar df&=0\; ,\\
2\, \bar P_{-}\,\bar de+d\bar f&=0 \; .
\end{split}
\end{equation}
In order to illustrate how to perform computations in this framework, let us explicitly derive the above equations. The differential $B_{1}$ written in terms of the single copy differentials $\d$ and $\bar\d$ acts on the components of the field $\Psi$ as
\begin{equation}
\begin{split}
    B_{1}(\Psi)=&(\d\otimes 1)\, f+(\d\otimes 1)\, e+(\d\otimes 1)\, \bar f\\
    +&(1\otimes\bar\d)\, f+(1\otimes\bar\d)\, e+(1\otimes\bar\d)\, \bar f\; .
\end{split}
\end{equation}
In the first line, notice that $f$ is an anti self-dual two-form in $dx$ and a zero-form in $d\bar x$ and, as a consequence, the action of $\d\otimes 1$ on this particular element vanishes. Indeed, $\d$ acting on the space of anti self-dual two-forms $K_{2}$ of the single copy kinematic algebra vanishes by degree: $(\d\otimes 1)f\equiv 0$. On the other hand, in the second term of the first line, $e$ is a one-form in $d x$ and a one-form in $d\bar x$, and hence the unbarred differential $\d$ acts on it as on an element of $K_{1}$ (or gauge field) of the unbarred copy of the kinematic algebra, namely $(\d\otimes 1)\, e=2\, P_{-}\, de$. Similarly, in the third term of the first line, $\bar f$ is a zero-form in $dx$ and an anti self-dual two-form in $d\bar x$. As a result, the unbarred differential $\d$ acts on it as an element of $K_{0}$ of the unbarred copy of the kinematic algebra, namely $(\d\otimes 1)\, \bar f=d\bar f$. Following the same reasoning for the second line of the above equation leads to
\begin{equation}
\begin{split}
    B_{1}(\Psi)=2\, P_{-}d\, e+d\bar f+\bar d f+2\, \bar P_{-}\bar d\, e\; .
\end{split}
\end{equation}
Notice that the first and the third terms are $(2,1)$-forms, whereas the second and fourth terms are $(1,2)$-forms. As a consequence, $B_{1}(\Psi)$ has two different components and $B_{1}(\Psi)=0$ implies the two independent field equations
\begin{equation}\label{firstoreoms22}
\begin{split}
2\, P_{-}\, d e+\bar df&=0\; ,\\
2\, \bar P_{-}\,\bar de+d\bar f&=0 \; .
\end{split}
\end{equation}
These field equations are invariant under linear gauge transformations, which can be obtained by acting with $B_{1}$ on gauge parameters $\Lambda$ using the same procedure described above, which leads to
\begin{equation}
\begin{split}
\delta^{(0)}e&=d\bar\lambda+\bar d\lambda=(\del_{\mu}\bar\lambda_{\bar\nu}+\bar\del_{\bar\nu}\lambda_{\mu})\, dx^{\mu}\otimes d\bar x^{\bar\nu}\;,\\
\delta^{(0)} f&=2\, P_{-}\, d\lambda=\tfrac{1}{2}\big\{-\del_{\mu}\lambda_{\nu}+\tfrac{1}{2}\, \epsilon_{\mu\nu\rho\sigma}\del^{\rho}\, \lambda^{\sigma}\big\}dx^{\mu}\wedge d x^{\nu}\; ,\\
\delta^{(0)} \bar f&=2\, \bar P_{-}\, \bar d\bar\lambda=\tfrac{1}{2}\big\{\bar \del_{\bar\mu}\bar\lambda_{\bar\nu}-\tfrac{1}{2}\, \bar\epsilon_{\bar\mu\bar\nu\bar\rho\bar\sigma}\bar \del^{\bar \rho}\, \bar\lambda^{\bar \sigma}\big\}d\bar x^{\bar \mu}\wedge d \bar x^{\bar\nu}\; .
\end{split}
\end{equation}

Notice that the field equations \eqref{firstoreoms} that follow from our double copy prescription are first order equations, whereas linearised self-dual gravity has second order differential operators acting on the graviton field. In order to obtain equations that can be more easily associated with self-dual gravity, we can eliminate $f$ and $\bar f$ from \eqref{firstoreoms} by acting with the de Rham differentials. More precisely, acting with a barred differential on the first equation in \eqref{firstoreoms} and acting with an unbarred differential on the second, leads to the second order equations
\begin{equation}\label{2ndeom}
    E:=2\, P_{-}d\, \bar d e=0\;,\;\;\; \bar E:=2\, \bar P_{-}d\, \bar d e=0\; ,
\end{equation}
which in components can be expressed as
\begin{equation}\label{secondorderdiff}
\begin{split}
E_{\mu\rho,\bar\nu\bar\sigma}:=\mathcal{R}_{\mu\rho,\bar\nu\bar\sigma}-\tfrac{1}{2}\, \epsilon_{\mu\rho}{}^{\lambda\tau}\, \mathcal{R}_{\lambda\tau,\bar\nu\bar\sigma}=0\; ,\\
\bar E_{\mu\rho,\bar\nu\bar\sigma}:=\mathcal{R}_{\mu\rho,\bar\nu\bar\sigma}-\tfrac{1}{2}\, \bar\epsilon_{\bar\nu\bar\sigma}{}^{\bar\lambda\bar\tau}\, \mathcal{R}_{\mu\nu,\bar\lambda\bar\tau}=0\; ,
\end{split}
\end{equation}
where $\mathcal{R}_{\mu\rho,\bar\nu\bar\sigma}$ is the following Riemann-like tensor
\begin{equation}
\mathcal{R}_{\mu\rho,\bar\nu\bar\sigma}:=-2\, \del_{[\mu}\bar\del_{[\bar\nu}e_{\rho]\bar\sigma]}\; .
\end{equation}
Notice that this object is anti-symmetric in each pair (barred and unbarred) of indices, but is not symmetric in the exchange of the two pairs of indices. Taking the supergravity solution of the strong constraint, where one identifies barred and unbarred coordinates and indices, the components of the field $e$ become $e_{\mu\nu}$ with a symmetric and an antisymmetric part. More precisely, we can split $e_{\mu\nu}$ as 
\begin{equation}\label{decompe}
    e_{\mu\nu}=h_{\mu\nu}+B_{\mu\nu}\; ,
\end{equation}
where the symmetric part $e_{(\mu\nu)}\equiv h_{\mu\nu}$ can be identified as the graviton, and the antisymmetrict part can be identified as a two-form or $B-$field: $e_{[\mu\nu]}\equiv B_{\mu\nu}$. From this decomposition  of the tensor fluctuation $e_{\mu\nu}$, one can identify the following linearized gauge transformations for the graviton and $B-$field
\begin{equation}
\begin{split}
\delta^{(0)}h_{\mu\nu}=\del_{\mu}\xi_{\nu}+\del_{\nu}\xi_{\mu}\;,\\
\delta^{(0)} B_{\mu\nu}=\del_{\mu}\zeta_{\nu}-\del_{\nu}\zeta_{\mu}\; ,
\end{split}
\end{equation}
where we introduced the gauge parameters $\xi$ and $\zeta$, which are linear combinations of $\lambda_{\mu}$ and $\bar\lambda_{\mu}$, namely $\xi_{\mu}=\tfrac{1}{2}\big(\lambda_{\mu}+\bar\lambda_{\mu}\big)$, $\zeta_\mu=\tfrac{1}{2}\big( \bar\lambda_{\mu}-\lambda_{\mu} \big)$. The above gauge transformations are, respectively, linearised diffeomorphisms and two-form gauge transformations. 

In the supergravity solution and with the decomposition of $e_{\mu\nu}$ into symmetric and antisymmetric parts as in \eqref{decompe}, the Riemann-like tensor can be expressed as 
\begin{equation}
\mathcal{R}_{\mu\nu,\rho\sigma}=R^{\rm lin.}_{\mu\nu\rho\sigma}+\del_{[\mu}H_{\nu]\rho\sigma}\; ,
\end{equation}
where $R^{\rm lin.}_{\mu\nu\rho\sigma}$ is the linearised Riemann tensor of general relativity and $H_{\mu\nu\rho}$ is the field-strength of the $B-$field given by $H_{\mu\nu\rho}=3\, \del_{[\mu}B_{\nu\rho]}$. 

We can extract the linearised self-duality equation for the graviton by taking a linear combination of the second order field equations \eqref{secondorderdiff} in the supergravity solution of the strong constraint. In particular, taking $\tfrac{1}{2}\big\{E_{\mu\rho,\nu\sigma}+\bar E_{\nu\sigma,\mu\rho}\big\}$ leads to
\begin{equation}
R^{\rm lin.}_{\mu\rho\nu\sigma}=\tfrac{1}{2}\, \epsilon_{\mu\rho}{}^{\lambda\tau}\, R^{\rm lin.}_{\lambda\tau\nu\sigma}\;,
\end{equation}
which is linearised self-dual gravity.
Taking the other linear combination $\tfrac{1}{2}\big\{E_{\mu\rho,\nu\sigma}-\bar E_{\nu\sigma,\mu\rho}\big\}$, on the other hand, yields 
\begin{equation}\label{minuscombi}
    \del_{[\mu}H_{\rho]\nu\sigma}=\tfrac{1}{2}\epsilon_{\mu\rho}{}^{\lambda\tau}\, \del_{\lambda}H_{\tau\nu\sigma}\; .
\end{equation}
We will now prove that the above equation for the $B-$field gives no propagating degrees of freedom. To this end, we take a trace of \eqref{minuscombi} to obtain
\begin{equation}
    \del^{\mu}H_{\mu\rho\sigma}=\epsilon_{\rho}{}^{\lambda\tau\mu}\, \del_{\lambda}H_{\tau\mu\sigma}\; .
\end{equation}
Using Hodge duality $H_{\mu\nu\rho}=\epsilon_{\mu\nu\rho\sigma}\, H^{\sigma}$, the above equation becomes
\begin{equation}
    \del^{\mu}H_{\mu\rho\sigma}=2\, \del_{\sigma}H_{\rho}\; ,
\end{equation}
which, in turn, implies that the symmetric part of the right-hand side vanishes:
\begin{equation}
    \del_{(\mu}H_{\nu)}=0\; .
\end{equation}
This is a Killing equation for $H_\mu$, whose solutions (constant or linear in $x$) are not propagating degrees of freedom, in the sense of finite superpositions of plane waves. 

Similarly, let us show that the anti-self-dual two-forms $f$ and $\bar f$ do not propagate further degrees of freedom. One can start from the second order equations \eqref{2ndeom} which, as we have discussed, propagate only one helicity of the graviton. These can be written as closure conditions:
\begin{equation}
\bar d\big(P_{-}d e\big)=0\;,\quad d\big(\bar P_{-}\bar d e\big)=0\;.
\end{equation}
Since $e$ is a $(1,1)$ form fluctuation on a topologically trivial background, the above equations can be integrated to \eqref{firstoreoms}, with $f$ and $\bar f$ playing the role of ``integration constants''. This means that, given a tensor $e$ satisfying \eqref{2ndeom}, one can find an $f$ and $\bar f$ so that \eqref{firstoreoms} are obeyed. More explicitly, upon identifying $x^\mu=\bar x^{\bar\mu}$, the equation involving $f$ reads  
\begin{equation}
    \del_{[\mu}e_{\nu]\rho}-\tfrac{1}{2}\epsilon_{\mu\nu}{}^{\lambda\tau}\, \del_{\lambda}e_{\tau\rho}+\tfrac{1}{2}\, \del_{\rho}f_{\mu\nu}=0\; .
\end{equation}
Looking for propagating degrees of freedom one can, for instance, use light-cone coordinates and assume that $\del_+$ is invertible. The $\rho=+$ component of the equation above then allows one to solve for $f_{\mu\nu}$ in terms of $e_{\mu\nu}$, thus showing that it does not carry independent degrees of freedom.

\section{Light-cone gauge and area preserving diffeomorphisms}
\label{Light-cone gauge and area preserving diffeomorphisms}

We will now make contact with the known kinematic algebra of self-dual Yang-Mills in light-cone gauge \cite{Monteiro:2011pc}. To do so, we will show that the $\BVB$ algebra discussed in the previous section reduces to a strict $\BV$ algebra upon taking light-cone gauge and partially solving the self-duality field equations. This $\BV$ algebra contains, in particular, the Lie algebra of area-preserving diffeomorphisms.

\subsection{Light-cone analysis}

We begin by introducing ``light-cone'' complex coordinates on Euclidean $\mathbb{R}^4$, defined as
\begin{equation}
x^\pm:=\frac{1}{\sqrt2}\,(z\pm it)\;,\quad w:=\frac{1}{\sqrt2}\,(x+iy)\;,\quad \bar w:=\frac{1}{\sqrt2}\,(x-iy)\;.    
\end{equation}
In these coordinates, the flat Euclidean metric reads
\begin{equation}
ds^2=2\,dx^+dx^-+2\,dwd\bar w\quad\rightarrow\quad\delta_{+-}=\delta^{+-}=1\;,\quad \delta_{w\bar w}=\delta^{w\bar w}=1\;,    
\end{equation}
and $\epsilon_{+-w\bar w}=1$. The non-trivial components of the self-duality equation of motion $(1-\star)F=0$ are given by
\begin{equation}\label{SDcomponents}
2\,F_{+w}=0\;,\quad F_{+-}+F_{w\bar w}=0\;,\quad 2\,F_{-\bar w}=0\;,   
\end{equation}
where we kept track of the factors of $2$ in order to match the normalization of the maps with the previous section. At this point, we impose light-cone gauge $A_+=0$. To do so we are assuming that $\del_+$ is invertible, which then ensures that there is no residual local symmetry. In light-cone gauge, the first equation in \eqref{SDcomponents} linearizes and we solve $F_{+w}=\del_+A_w=0$. Since $\del_+$ can be inverted, this sets $A_w=0$, thereby eliminating one helicity of the gluon. We are thus left with two non-vanishing components of the gauge field: $A_\mu=(0,A_-,0,A_{\bar w})$. One could now solve explicitly the second equation in \eqref{SDcomponents} in terms of a single degree of freedom, but we prefer not to do so at this stage, and instead rewrite the remaining two equations in a 2D covariant form. To this end, we define 
\begin{equation}
z^\alpha:=(x^+,w)\;,\quad\del_\alpha=(\del_+,\del_w)\;,\quad\tilde\del^\alpha:=(\del^+,\del^{w})=(\del_-,\del_{\bar w})\;,
\end{equation}
with $\alpha,\beta,\cdots$ indices taking values $(+,w)$. Notice that there is no 2D metric to raise and lower $\alpha$ indices, and the four-dimensional Laplacian is given by $\B=2\,\tilde\del^\alpha\del_\alpha$.
The surviving components of the gauge field can then be written as a two-dimensional vector: $A^\alpha=(A^+,A^w)=(A_-,A_{\bar w})$, and the two remaining equations in \eqref{SDcomponents} can be written as
\begin{equation}\label{2Dform}
\begin{split}
F_{+-}+F_{w\bar w}&\quad\longrightarrow\quad\del_\alpha A^\alpha\;\\    
F_{-\bar w}&\quad\longrightarrow\quad\,\tilde\del^\alpha A^\beta-\tilde\del^\beta A^\alpha+[A^\alpha,A^\beta]\;,
\end{split}    
\end{equation}
with square brackets denoting the color Lie algebra commutator. Notice that the second equation above is a zero curvature condition in two dimensions, despite the fact that the fields depend on four coordinates, which makes it formally similar to a Chern-Simons theory in Lorentz gauge\footnote{For a similar observation in twistor space see e.g. \cite{Borsten:2022vtg,Bittleston:2020hfv}.}. In order to study the kinematic algebra associated to \eqref{2Dform}, we will consider $\del_\alpha A^\alpha=0$ as a constraint on $A^\alpha$, while we will not impose the dynamical equation.

\subsection{Light-cone kinematic algebra}

As we have explained in \autoref{Sec:Loo}, the full self-dual theory is encoded by a (strict) $L_\infty$ algebra. The same is true for the above system of field equations \eqref{2Dform} without gauge symmetry. This is obtained from the gauge invariant and completely off-shell one described in section \ref{Sec:Loo} upon gauge fixing and partially solving the equations. Stripping off color as done in \autoref{Sec:BVoo} one obtains the corresponding differential graded commutative algebra. Since we have gauge fixed the theory completely, the space $K_0$ associated to gauge parameters trivializes. The space $K_1$ of color-stripped gauge fields reduces to the space of 2D transverse vectors $\cA^\alpha$ (as before, we use a different font to recall that $\cA^\alpha$ has no color degrees of freedom). Finally, the space $K_2$ reduces to 2D bivectors $\cF^{\alpha\beta}$, associated to the equation of motion $F^{\alpha\beta}$ in \eqref{2Dform}. This can be visualized in the following diagram:
\begin{equation}\label{lcKdiagram}
\begin{tikzcd}[row sep=4mm]
0\arrow[r]& K^{\rm l.c.}_1 \arrow[r, "\d "]
&\arrow[l, bend left=50, "b"] K_2^{\rm l.c.} \\
& \cA^\alpha & \cF^{\alpha\beta}
\end{tikzcd}  \;,  
\end{equation}
where $\cA^\alpha$ is subject to $\del_\alpha\cA^\alpha=0$, and one can see that the degree in $\cK^{\rm l.c.}$ coincides with the ``polyvector degree''. The differential $\d $ and $b$ operator act as
\begin{equation}\label{QbLC}
\big(\d \cA\big)^{\alpha\beta}=2\,\tilde\del^\alpha\cA^\beta-2\,\tilde\del^\beta\cA^\alpha\;,\quad\big(b\cF\big)^\alpha=\del_\beta\cF^{\beta\alpha}\;,    
\end{equation}
with $b\cA\equiv0$ and $\d \cF\equiv 0$. One can check that they obey $\d b+b\d =\B$ by using the constraint $\del_\alpha\cA^\alpha=0$ and the Schouten identity in two dimensions $\tilde\del^{[\alpha}\cF^{\beta\gamma]}\equiv0$. By degree, the only non-vanishing product $m_2$ is between vectors and is obtained by color stripping the nonlinear term in the equation \eqref{2Dform}:
\begin{equation}\label{m2LC}
m^{\alpha\beta}_2(\cA_1,\cA_2)=2\,\cA_1^\alpha\,\cA_2^\beta-2\,\cA_1^\beta\,\cA_2^\alpha\;.    
\end{equation}
The differential $\d$, operator $b$ and product $m_2$ follow from the gauge invariant ones of the off-shell theory in \autoref{Sec:BVoo} upon setting $\lambda, \cA_+,\cA_w$ and $\cE_{+w}$ to zero.

We now repeat the procedure of the previous section, and define the kinematic bracket $b_2$ by the failure of $b$ to obey the Leibniz rule with respect to $m_2$:
\begin{equation}
b_2(u_1,u_2)=bm_2(u_1,u_2)-m_2(bu_1,u_2)-(-1)^{u_1}m_2(u_1,bu_2)\;.    
\end{equation}
Using the expressions \eqref{QbLC} and \eqref{m2LC} one finds two non-vanishing brackets:
\begin{equation}\label{b2LC}
\begin{split}
b^\alpha_2(\cA_1,\cA_2)&=\del_\beta\,m^{\beta\alpha}_2(\cA_1,\cA_2)=2\,\big[\cA_1,\cA_2\big]^\alpha_{\rm SN}\;,\\
b^{\alpha\beta}_2(\cA,\cF)&=m^{\alpha\beta}_2(\cA,b\cF)=2\,\big[\cA,\cF\big]^{\alpha\beta}_{\rm SN}\;,
\end{split}    
\end{equation}
where $[\,,\,]_{\rm SN}$ denotes the Schouten-Nijenhuis bracket of polyvectors. In the case at hand, both brackets can be written in terms of standard Lie derivatives:
\begin{equation}
\begin{split}
\big[\cA_1,\cA_2\big]^\alpha_{\rm SN}&=\cA_1^\beta\del_\beta\cA_2^\alpha-\cA_2^\beta\del_\beta\cA_1^\alpha=\cL_{\cA_1}\cA_2^\alpha\;,\\
\big[\cA,\cF\big]^{\alpha\beta}_{\rm SN}&=\cA^\gamma\del_\gamma\cF^{\alpha\beta}+2\,\cF^{\gamma[\alpha}\del_\gamma\cA^{\beta]}=\cL_\cA\cF^{\alpha\beta}\;.
\end{split}    
\end{equation}
This shows that $b_2$ generates the algebra of area-preserving diffeomorphisms when restricted to vectors, and extends to a graded Lie algebra on the space of polyvector fields via the Schouten-Nijenhuis bracket.
This is not a mere change of perspective: the bracket $b_2$ is compatible with the product $m_2$ in the graded Poisson sense:
\begin{equation}
b_2\big(\cA_1,m_2(\cA_2,\cA_3)\big)=m_2\big(b_2(\cA_1,\cA_2),\cA_3\big)+m_2\big(\cA_2,b_2(\cA_1,\cA_3)\big)\;,  
\end{equation}
which goes beyond the properties of Lie algebras. In fact, the triplet $(b,m_2,b_2)$ forms a BV algebra, which gets deformed to $\BV$ upon including the differential $\d $, exactly as in Chern-Simons theory.

To make contact with the off-shell and gauge independent algebra presented in \autoref{Sec:BVoo}, notice that the brackets \eqref{b2LC} coincide with the off-shell ones \eqref{b2} upon setting $\lambda, \cA_+, \cA_w$ and $\cE_{+w}$ to zero and using the constraint $\del_\alpha\cA^\alpha=0$. Consistently, the homotopy maps $\theta_3$ in \eqref{theta3} vanish in this case. This is easily seen from the fact that both $\theta_3(\cA_1,\cA_2,\cA_3)$ and $\theta_3(\cE,\cA_1,\cA_2)$ contain a three-form, which vanishes identically in the present effective 2D setting. For instance, the homotopy $\theta_3(\cA_1,\cA_2,\cA_3)$
in components reads
\begin{equation}\label{thetaex}
\theta_{3\mu}(\cA_1,\cA_2,\cA_3)=-\epsilon_{\mu\nu\rho\sigma}\,\cA_1^\nu\cA_2^\rho\cA_3^\sigma \;. 
\end{equation}
As we have discussed, after taking light-cone gauge and eliminating one helicity $\cA^\mu_i$ reduce to $\cA_i^\alpha$, with $\alpha=(+,w)$ the two dimensional index.
This makes \eqref{thetaex} vanish, due to $\cA_1^{[\alpha}\cA_2^\beta\cA_3^{\gamma]}\equiv0$ in two dimensions. At this point it is worth emphasizing 
that the self-duality condition has a dramatic algebraic outcome: fixing the light-cone gauge alone does \emph{not} reduce the $\BVB$ algebra to a strict one. It is only after setting to zero the helicity $\cA^{\bar w}$ that all homotopies vanish and one obtains a $\BV$ algebra. 

We conclude this section by making contact with the description of the light-cone kinematic algebra in terms of a single ``scalar'' degree of freedom. To do so, we introduce the Schouten-Nijenhuis bracket between a vector and a scalar, as well as a bivector and a scalar, although they are not part of $b_2$:
\begin{equation}
\big[\cA,\chi\big]_{\rm SN}=\cA^{\beta}\del_\beta\chi\;,\quad \big[\cF,\chi\big]^\alpha_{\rm SN}=\cF^{\beta\alpha}\del_\beta\chi \;.  
\end{equation}
With this, one can solve the constraint $\del_\alpha\cA^\alpha=0$ explicitly using the two-dimensional $\epsilon^{\alpha\beta}$ symbol, normalised as $\epsilon^{w+}=1$, viewed as a special bivector:
\begin{equation}\label{phi}
\cA^\alpha=-\big[\epsilon,\Phi\big]^\alpha_{\rm SN}=\epsilon^{\alpha\beta}\del_\beta\Phi\;.    
\end{equation}
Since in this light-cone treatment we assume $\del_+$ to be invertible, we can express $\Phi=\frac{1}{\del_+}\cA^w$, which shows that $\Phi$ is \emph{not} a standard four-dimensional scalar field.
Using the solution \eqref{phi}, the product $m_2$ between vectors yields
\begin{equation}
m_2^{\alpha\beta}(\cA_1,\cA_2)=2\,\epsilon^{\alpha\beta}\big\{\Phi_1,\Phi_2\big\} \;,   
\end{equation}
in terms of the Poisson bracket
\begin{equation}
\big\{\Phi_1,\Phi_2\big\}:=\big[\big[\epsilon,\Phi_1\big]_{\rm SN},\Phi_2\big]_{\rm SN}=\epsilon^{\alpha\beta}\del_\alpha\Phi_1\del_\beta\Phi_2\;,    
\end{equation}
while the Lie bracket $b_2$ reduces to
\begin{equation}
b_2^\alpha(\cA_1,\cA_2)=2\,\big[\epsilon,\big\{\Phi_1,\Phi_2\big\}\big]_{\rm SN}^\alpha=-2\,\epsilon^{\alpha\beta}\del_\beta\big\{\Phi_1,\Phi_2\big\}\;.    
\end{equation}

\section{Light-cone double copy to self-dual gravity}
\label{Double Copy and Self-Dual Gravity}

Since in \autoref{Sec:BVoo} we limited ourselves to the double copy of the free theory, here we take advantage of the fact that the light-cone kinematic algebra described in \autoref{Light-cone gauge and area preserving diffeomorphisms} is strict, in order to perform an exact double copy, following again the prescription of \cite{Bonezzi:2022bse}.
We do so in order to strengthen the evidence that these $\BVB$ algebras are the structures responsible for consistency of the double copy procedure. Our starting point is thus the $\BV$ algebra of SDYM in light-cone gauge, described by the graded vector space $\cK^{\rm l.c.}$ \eqref{lcKdiagram}, with differential and $b$ operator \eqref{QbLC}, product \eqref{m2LC} and kinematic bracket \eqref{b2LC}.

As we have discussed in \autoref{Sec:BVoo}, we shall take the tensor product of two copies of $\cK^{\rm l.c.}$, and use the strong constraint \eqref{strong} to eventually identify the doubled coordinates. The graded vector space $\mathcal{X}^{\rm{DC}}$ that results from the tensor product contains three vector spaces:
\begin{equation}
    \mathcal{X}^{\rm{DC}}=\bigoplus_{i=0}^{2}X^{\rm{DC}}_{i}\; ,
\end{equation}
where the degree in $\cX^{\rm DC}$ is defined by $|\Psi|=|u|+|\bar u|-2$ for an element $\Psi=u\otimes\bar u$, and the shift by two is to make contact with standard $L_\infty$  grading for the double copy theory.
The space $X^{\rm{DC}}_{0}=K^{\rm l.c.}_{1}\otimes \bar K^{\rm l.c.}_{1}$ is then identified as the space of gravity fields $h^{\alpha\bar\beta}$ with $\alpha$ and $\bar\beta$  the two dimensional indices introduced in the previous section associated to $\mathcal{K}^{\rm l.c.}$ and $\bar{\mathcal{K}}^{\rm l.c.}$, respectively. The space $X^{\rm{DC}}_{1}=K^{\rm l.c.}_{1}\otimes \bar K_{2}^{\rm l.c.}\oplus K^{\rm l.c.}_{2}\otimes \bar K_{1}^{\rm l.c.}$ is the two-component space of field equations $\cE^{\alpha\beta\bar\gamma}$ and $\cE^{\bar\alpha\bar\beta\gamma}$, and the final space $X^{\rm{DC}}_{2}=K^{\rm l.c.}_{2}\otimes \bar K_{2}^{\rm l.c.}$ is a space of identities \footnote{Even though there is no gauge symmetry, the theory still obeys a set of identities, as we will show momentarily.} $\cN^{\alpha\beta\bar\gamma\bar\delta}$. The gravity field $h^{\alpha\bar\beta}$ has no symmetry between the indices, while both equations $\cE^{\alpha\beta\bar\gamma}$, $\cE^{\bar\alpha\bar\beta\gamma}$, and identities $\cN^{\alpha\beta\bar\gamma\bar\delta}$ are antisymmetric in the couples of (un)barred indices. Similarly to the construction of the vector $\cA^\alpha$ from $A_\mu$ by taking light-cone gauge and partially solving the self-duality relations, $h^{\alpha\bar\beta}$ can be thought as coming from gauge fixing the graviton $h_{\mu\nu}$ in light-cone gauge and partially solving the gravity self-duality field equations.
The graded vector space above forms a chain complex as: 
\begin{equation}
\begin{tikzcd}[row sep=2mm]
X^{\rm{DC}}_{0}\arrow{r}{B_{1}} & X^{\rm{DC}}_{1}\arrow{r}{B_{1}} & X^{\rm{DC}}_{2} \\
h^{\alpha\bar\beta}& \begin{matrix}\cE^{\alpha\beta\bar\gamma}\\[4mm]
\bar{\cE}^{\bar\alpha\bar\beta\gamma}\end{matrix}&\mathcal{N}^{\alpha\beta\bar\gamma\bar\delta} \;,
\end{tikzcd}    
\end{equation}
where $B_1$ is the differential of the double copy theory to be identified in the following. 

The next step in the double copy prescription is indeed to construct the multilinear maps of the $L_\infty$ algebra of the gravity theory using the kinematic maps of the two copies of SDYM. Given that the kinematic algebra of light-cone gauge SDYM is a strict BV$^{\square}$, in the gravity theory there only exist a differential $B_{1}$ and a two-bracket $B_{2}$. These are given in terms of the SDYM maps by 
\begin{equation}\label{Bn}
\begin{split}
        B_{1}&=\d \otimes 1+1\otimes \bar \d \; ,\\
        B_{2}&=\tfrac{1}{2}\, b_{2}\otimes \bar m_{2}-\tfrac{1}{2}\, m_{2}\otimes \bar b_{2}\; ,
\end{split}
\end{equation}
following the general procedure of \cite{Bonezzi:2022bse}. Here $\d$, $m_{2}$ and $b_{2}$ act on $\mathcal{K}^{\rm l.c.}$, while $\bar\d$, $\bar m_{2}$ and $\bar b_{2}$ act on $\bar{\mathcal{K}}^{\rm l.c.}$.

The gravity maps $B_{1}$ and $B_{2}$ obey the $L_{\infty}$ relations \eqref{linftyrels1} without higher brackets: $B_n=0$ for $n>2$. This follows, according to the general discussion in \cite{Bonezzi:2022bse,Bonezzi:2023ced}, from the fact that $\d$, $m_{2}$ and $b_{2}$ (and their barred counterparts) obey the BV$^{\square}$ relations without higher maps. Intuitively, $B_2$ obeys the Jacobi identity because both $b_2$ and $\bar b_2$ are Lie brackets and $m_2$, $\bar m_2$ are both associative products.
Furthermore, the Leibniz property of $B_1$ with respect to $B_2$ in ensured by the strong constraint \eqref{strong}: indeed, identifying $\B\equiv\bar\B$ removes the $\B-\bar\B$ deformation arising from the two copies of \eqref{DefLeibniz}. Importantly, since there are no higher $L_\infty$ brackets $B_n$, the gravitational equations obtained from \eqref{Bn} are exact. 

We now  provide the explicit action of the gravitational differential and bracket. Starting from $B_1$, acting on the field it produces two components:
\begin{equation}
\begin{split}
B_{1}^{\alpha\beta\bar\gamma}\big(h\big)=2\, \tilde\del^{\alpha}h^{\beta\bar\gamma}-2\, \tilde\del^{\beta}h^{\alpha\bar\gamma}\; ,\\
B^{\bar\alpha\bar\beta\gamma}_{1}\big(h\big)=-2\, \tilde\del^{\bar\alpha}h^{\gamma\bar\beta}+2\, \tilde\del^{\bar\beta}h^{\gamma\bar\alpha}\; ,
\end{split}
\end{equation}
while the action  on the space of equations reads
 \begin{equation}
     B^{\alpha\beta\bar\gamma\bar\delta}_{1}(\cE)=4\, \Big\{\tilde\del^{\bar\delta}\cE^{\alpha\beta\bar\gamma}-\tilde\del^{\bar\gamma}\cE^{\alpha\beta\bar\delta}+\tilde\del^{\alpha}\cE^{\bar\delta\bar\gamma\beta}-\tilde\del^{\beta}\cE^{\bar\delta\bar\gamma\alpha}\Big\}\; ,
 \end{equation}
which justifies the existence of the space of identities $\cN^{\alpha\beta\bar\gamma\bar\delta}$, since nilpotency of the differential gives
\begin{equation}
 \big(B_1^2h\big)^{\alpha\beta\bar\gamma\bar\delta}=4\, \Big\{\tilde\del^{\bar\delta}B_{1}^{\alpha\beta\bar\gamma}(h)-\tilde\del^{\bar\gamma}B_{1}^{\alpha\beta\bar\delta}(h)+\tilde\del^{\alpha}B_{1}^{\bar\delta\bar\gamma\beta}(h)-\tilde\del^{\beta}B_{1}^{\bar\delta\bar\gamma\alpha}(h)\Big\}\equiv0\; ,
\end{equation}
and the nonlinear completion is encoded in $B_{2}(h,\cE)$ which we do not display. 

The interactions of the theory are governed by the action of $B_{2}$ on two fields
\begin{equation}
\begin{split}
    B^{\alpha\beta\bar\gamma}_{2}(h_{1},h_{2})&=4\, \big[h_{1}^{\alpha},h_{2}^{\beta}\big]^{\bar\gamma}_{\overline{\rm{SN}}}=4\,\Big\{ h_{1}^{\alpha\bar\delta}\del_{\bar\delta}h_{2}^{\beta\bar\gamma}-h_{2}^{\beta\bar\delta}\del_{\bar\delta}h_{2}^{\alpha\bar\gamma} \Big\}\\
    B^{\bar\alpha\bar\beta\gamma}_{2}(h_{1},h_{2})&=-4\, \big[h_{1}^{\bar\alpha},h_{2}^{\bar\beta}\big]^{\gamma}_{\rm{SN}}=-4\, \Big\{ h_{1}^{\delta\bar\alpha}\del_{\delta}h_{2}^{\gamma\bar\beta}-h_{2}^{\delta\bar\beta}\del_{\delta}h_{1}^{\gamma\bar\alpha} \Big\}\; ,
\end{split}
\end{equation}
where the notation $\big[h_{1}^{\bar\alpha},h_{2}^{\bar\beta}\big]^{\gamma}_{\rm{SN}}$ means to take the Schouten-Nijenhuis bracket by viewing $h^{\alpha\bar\beta}$ as a vector $\big(\cV^\alpha\big)^{\bar\beta}$, with the $\bar\beta$ index as a spectator, and vice versa for the other component.
Thus we can construct the field equations of the gravity theory in $L_{\infty}$ form
\begin{equation}
    B_{1}(h)+\tfrac{1}{2}B_{2}(h,h)=0\; ,
\end{equation}
which in components read
\begin{equation}\label{eomsh}
\begin{split}
2\, \tilde\del^{\alpha}h^{\beta\bar\gamma}-2\, \tilde\del^{\beta}h^{\alpha\bar\gamma}+2\, \big[h^{\alpha},h^{\beta}\big]^{\bar\gamma}_{\overline{\rm{SN}}}=0\; ,\\
-2\, \tilde\del^{\bar\alpha}h^{\gamma\bar\beta}+2\, \tilde\del^{\bar\beta}h^{\gamma\bar\alpha}-2\, \big[h^{\bar\alpha},h^{\bar\beta}\big]^{\gamma}_{\rm{SN}}=0\; .
\end{split}
\end{equation}
We now turn to proving that these first order equations indeed describe self-dual gravity. First, the gravity field $h^{\alpha\bar\beta}$ is subject to the following constraints
\begin{equation}\label{consonh}
    \del_{\alpha}h^{\alpha\bar\beta}=0\; ,\quad
    \del_{\bar\beta}h^{\alpha\bar\beta}=0\; ,
\end{equation}
which follow from the constraint $\del_{\alpha}\cA^{\alpha}=0$. The constraints \ref{consonh} can be solved by writing the gravity field as
\begin{equation}\label{gravphi}
    h^{\alpha\bar\beta}=\epsilon^{\alpha\gamma}\, \epsilon^{\bar\beta\bar\delta}\, \del_{\gamma}\, \del_{\bar\delta}\, \phi\; ,
\end{equation}
where $\phi$ is a ``scalar'' degree of freedom carrying the single helicity $+2$ of the self-dual graviton. We can take the first field equation in \eqref{eomsh}, which in terms of $\phi$ reads
\begin{equation}\label{eqphi}
    -\epsilon^{\bar\gamma\bar\delta}\del_{\bar\delta}\Big\{ \epsilon^{\alpha\beta}\, \big(\Box \phi+\epsilon^{\bar\alpha\bar\beta}\, \epsilon^{\gamma\delta}\, \del_{\bar\alpha}\del_{\gamma}\phi\, \del_{\bar\beta}\del_{\delta}\phi\big) \Big\}=0\; .
\end{equation}
In order to make contact with the results by Monteiro and O'Connell in \cite{Monteiro:2011pc}, we solve the strong constraint by choosing the supergravity solution $\del_{\alpha}=\del_{\bar\alpha}$. The form of $h^{\alpha\bar\beta}$ in \eqref{gravphi} then coincides with the self-dual graviton and, in turn, \eqref{eqphi} implies the Plebanski equation of self-dual gravity:
\begin{equation}\label{plebanski}
    \Box\phi+2\, \del^{2}_{w}\phi\, \del^{2}_{+}\phi-2\, \big(\del_{+}\del_{w}\phi\big)^{2}=0\; .
\end{equation}
Notice that in removing the derivative operator $\epsilon^{\bar\gamma\bar\delta}\del_{\bar\delta}$ to write the Plebanski equation there is no loss of information, since one can remove an invertible $\del_+$ to obtain \eqref{plebanski}.

\section{Conclusions and outlook}
\label{Conclusions and Outlook}

We showed how the framework of homotopy algebras can help to give a direct construction of a fully gauge independent kinematic algebra in the self-dual sector of YM theory, thus generalising previous results from the literature. We also demonstrated that the double copy procedure finds a natural generalisation in this language. 

Doubtless the elegant structures arising in this context are partly due to the simplicity of the self-dual sector, rooted in its integrability. However, as mentioned in the introduction, there is a notion of perturbation around this sector. It would be interesting to see if our results can be extended in this way, and whether they reproduce a gauge-independent version of \cite{Chen:2019ywi,Chen:2021chy,Ben-Shahar:2022ixa}. More generally, this gives a promising alternative perturbation scheme, different from the post-Minkowskian and post-Newtonian frameworks, which could perhaps teach us new lessons about color-kinematics duality and the double copy.

On the other hand, the self-dual sector is valuable in its own right as a toy model in which to address difficult questions pertaining to the full theory.
In particular, the kinematic algebra constructed in this paper shows that self-dual Yang-Mills is a promising midpoint between the simplest Chern-Simons case and the much more complex kinematic algebra of Yang-Mills theory. Due to its relative simplicity, we hope to be able to bootstrap the results of this paper to a complete characterization of the SDYM kinematic algebra to all orders. This could give valuable insights into the much more challenging case of full Yang-Mills.  

Another potential application is in the study of double field theory (DFT) \cite{Hull:2009mi,Hohm:2010jy,Hohm:2010pp}. DFT is a natural intermediate product in the double copy algorithm developed in \cite{Bonezzi:2022bse,Bonezzi:2022yuh,Diaz-Jaramillo:2021wtl}, which we have also followed here. We intend to use this to construct a self-dual sector of DFT that goes beyond self-dual gravity. One could then address difficult questions relating to the geometric formulation of DFT, initially in this simplified set-up, and subsequently in the full theory, recast as a perturbation around this sector.

\subsection*{Acknowledgements}
We would like to thank Christoph Chiaffrino, Olaf Hohm, Eric Lescano and Christian Saemann for helpful discussions.

F.DJ. thanks Durham University for hospitality during the early stages of this project. The work of F.DJ. is supported by the Deutsche Forschungsgemeinschaft (DFG, German Research Foundation) - Projektnummer 417533893/GRK2575 ``Rethinking Quantum Field Theory". The work of R.B. is supported by the European Research Council (ERC) under the European Union’s Horizon 2020 research and innovation programme (grant agreement No 771862). S.N. is supported in part by STFC consolidated grant T000708.

\providecommand{\href}[2]{#2}\begingroup\raggedright\endgroup

\end{document}